\documentclass[aps,utf8,floatfix,twocolumn,final,prl,showpacs]{revtex4-2}

\usepackage{epsfig}
\usepackage{graphicx}
\usepackage[ansinew,utf8]{inputenc}
\usepackage{array}
\usepackage{gensymb}
\usepackage{braket}
\usepackage{hyperref}
\usepackage{tabularx}
\newcolumntype{C}[1]{>{\centering\let\newline\\\arraybackslash\hspace{0pt}}m{#1}}
\usepackage{placeins}

\usepackage{xcolor}

\usepackage{lineno}

\usepackage{gensymb}

\usepackage{textcomp}

\begin{document}

\preprint{APS/123-QED}

\title{From Bulk to Nanowire: A Dimensional Journey with CeIn$_3$}

\author{M. H. Carvalho,$^{1,2}$, D. Zau$^1$, A. P. Reyes{$^{3}$}, R. Cong{$^{3}$}, S. D. House{$^{4}$}, H. P. Pizzi{$^{1}$}, A. M. Caffer{$^{1}$}, D.S. Passos{$^{1}$}, R. C. Santos{$^{1}$}, G. S. Freitas{$^{2}$}, K. R. Pirota{$^{1}$}, R. R. Urbano{$^{1}$}, P. G. Pagliuso$^{1,2}$}

\affiliation{$^1$ Instituto de F\'isica Gleb Wataghin, UNICAMP, Campinas-SP, 13083-859, Brazil. \\ $^2$ Los Alamos National Laboratory, Los Alamos, NM 87545, USA. \\ $^3$ National High Magnetic Field Laboratory, Tallahassee, Florida 32310, USA\\ $^4$ Center for Integrated Nanotechnologies (CINT), Sandia National Laboratories, New Mexico 87185, USA.}

\date{\today}

\begin{abstract}
In this work, we have explored the Metallic-Flux Nanonucleation method to synthesize single crystals and nanowires (diameter $\approx$ 170 nm) of CeIn$_{3}$ and compare their properties. The effects of reduced dimensionality were systematically investigated using Energy Dispersive Spectroscopy (EDS), Selected area electron diffraction (SAED), magnetic susceptibility, heat capacity, and Nuclear Magnetic Resonance (NMR). Semi-quantitative EDS analysis revealed a Ce:In ratio of 1:3.1(1), and the SAED results confirmed that the nanowires are polycrystalline with a cubic unit cell. Magnetic susceptibility, specific heat, and NMR data indicated a suppression of the antiferromagnetic transition to $T_N$ $\approx$ 2.4 K compared to the bulk value ($\approx$ 10 K). Furthermore, NMR analysis at temperatures below 2.8 K showed a reduced quadrupole frequency, $\nu_Q$ $\approx$ 1.77(2)  MHz, and provided evidence of polycrystalline nanowires formed within the nanoporous alumina template, in agreement with SAED results. We attribute these findings to an increasing magnetic order frustration induced by dimensionality in CeIn$_{3}$ nanowires.  
\end{abstract}

\maketitle


\section{\label{sec:level1}INTRODUCTION}
Interacting many-body systems in crystalline solids generally obey the symmetry properties of periodic lattices. Their microscopic description is governed by translational symmetry in three dimensions, which enables the application of Bloch's theorem, leading to the formation of energy bands that define the electronic structure of solids \cite{ashcroft1976solid}. Interactions within these systems give rise to the phenomenon of spontaneous symmetry breaking, manifesting in forms such as magnetic and crystalline order, and superconductivity \cite{fetter2003quantum}. A platform for exploring these properties is provided by reducing the dimensionality of the material. In particular, when one of the dimensions of the system becomes comparable to a fundamental length scale, the disruption of translational symmetry can alter its ground state. Furthermore, in reduced spatial dimensions, many-body correlation effects become more pronounced, as Coulomb interactions between electrons play an increasingly dominant role in the behavior of the material.

In Ce-based intermetallic heavy fermions (HF) materials, \textit{f} electrons hybridize with conduction electrons, and the Ruderman--Kittel--Kasuya--Yosida (RKKY) interaction, which favors long-range magnetic order, typically competes with the Kondo effect. As a result of this competition, a tuning of a magnetically ordered state toward a quantum phase transition at $T$ = 0 may occur. The balance of both interactions can be tuned by external parameters such as doping \cite{berry2010magnetic}, magnetic field \cite{ebihara2004emergent,moll2017emergent}, applied pressure \cite{mathur1998magnetically,grosche2001superconductivity}, and, of interest here, reduced dimensionality\cite{shishido2010tuning}. Unconventional superconductivity (USC) and Non-Fermi liquid behavior often occur in the vicinity of such quantum critical points \cite{paschen2021quantum}. 

The comparative study on the series Ce$_{m}M_{n}$ In$_{3m+2n}$ ($M =$ Co, Rh, Ir; $n = 0, 1$; $m = 1, 2$) previously allowed the investigation of the role of dimensionality in these compounds even in a single crystal form. For $m$ = 1 and $n$ = 0, one has the bulk three-dimensional (3D) cubic compound CeIn$_{3}$ that presents an antiferromagnetic (AFM) ordering,  $T_N \approx$ 10 K at ambient pressure, and undergoes a transition to a superconducting state ($T_C$ = 0.2 K) at critical pressure $P\approx$ 25 kbar \cite{mathur1998magnetically,grosche2001superconductivity}. For $m$ = 1 and $n$ = 1, one finds the more two-dimensional (2D) tetragonal compounds Ce$M$In$_{5}$, so-called 115, which present alternating layers of Ce-In$_3$ and $M$-In$_2$ along the $c$ axis \cite{d2011progress}. Depending on the transition metal element $M$, the 115 compounds can be unconventional superconductor, e.g. $M$ = Co, $T_C$ = 2.3 K or $M$ = Ir, {($T_C \approx 0.4$ K)},\cite{petrovic1,petrovic2} or antiferromagnetic (AFM) $M$ = Rh, with $T_N$ = 3.8 K, at atmospheric pressure, and becoming superconducting only under applied pressure \cite{d2011progress}. From these studies, it has been argued that unconventional superconductivity generated by magnetic fluctuations tends to be favored in the 2D 1-1-5 families, while AFM is favored for 3D bulk materials \cite{d2011progress}.

Rare earth metal (RE)In$_3$ compounds have already been extensively studied in the literature \cite{buschow1969,amara1994,amara1995,pagliuso2006,moll2017,simeth2023}. However, these compounds have been little explored when it comes to tuning their dimensionality in a controllable way, from 3D to 2D and 1D. One of the pioneering works in this direction reported the growth by molecular beam epitaxy of 2D CeIn$_{3}$/LaIn$_{3}$ heterostructures. The suppression of the magnetic order of CeIn$_{3}$ was observed with the reduction of the thickness of the CeIn$_{3}$ layers and the effective mass of the electron was further enhanced. In addition, the 2D confined heterostructures showed low-temperature deviations of the electronic properties, which were associated with the dimensional tuning of quantum criticality \cite{shishido2010tuning}.
Furthermore, Rosa et al. performed a study on the effect of dimensionality on the physical properties of the intermetallic compound GdIn$_{3}$ using the Metallic-Flux Nanonucleation (MFNN) method to obtain this compound in nanowire form. Magnetic susceptibility and specific heat measurements showed a drastic suppression of the AFM order temperature (T$^{3D}_{N} = 45$ K) of the single crystal bulk to a T$^{LD}_{N} = 3.8$ K for the nanowire system (diameter d $\approx 200$  and length l $\approx 30$ $\mu$m). Since Gd$^{3+}$ is a pure spin $S$ = 7/2 ion ($L$ = 0), such reduction was associated with changes in the RKKY magnetic interaction \cite{ROSA201414}.
The cubic intermetallic HF compound CeIn$_{3}$ is an excellent compound to understand, and ultimately control, quantum phenomena in \textit{f}-based low-dimensional systems, as it presents a variety of interesting physical properties, such as RKKY magnetic interaction, Kondo effect, crystalline electric field (CEF), non-Fermi-liquid behavior and the interaction between antiferromagnetism and unconventional superconductivity. However, the growth of intermetallic nanowires containing a rare-earth element has been
challenging \cite{ROSA201414, Carvalho_2022, tang2021, zhang2022, gou2025,moura2016,moura2017,alexsandro2021,cruz2022manganese,Raul}.

After intense efforts\cite{Carvalho_2022}, we have established a successful route to synthesize CeIn$_{3}$ nanowires with diameter of 170(5) nm by MFNN method.  The crystal structure and stoichiometry of the nanowires were investigated by Energy Dispersive X-ray spectroscopy (EDS) and Selected Area Electron Diffraction (SAED). In addition, the temperature dependence of the magnetic susceptibility and specific heat of an Al$_{2}$O$_{3}$ template with the CeIn$_{3}$ nanowires was determined the nanowires macroscopic properties. Importantly, to gain microscopic insights about the properties of the CeIn$_{3}$ nanowires, we have employed the Nuclear Magnetic Resonance (NMR) technique to investigate the microscopic nature of the relevant interactions in the nanowire system. 

Our results reveal a drastic suppression of the antiferromagnetic
transition from the bulk (T$^{3D}_{N} = 10 $ K) to the CeIn$_{3}$ nanowire systems (T$^{LD}_{N} = 2.3$ K) as similarly observed in their GdIn$_{3}$ counterparts \cite{ROSA201414}.  In particular, the NMR results reported here clearly indicate a disordered AFM state below $T_N$  arising from an arrangement of polycrystalline  CeIn$_3$ nanowires. 

We discuss this suppression of $T_N$ due to an increase of frustration of the magnetic order induced by reducing dimensionality, taking into account how the relevant interactions such as RKKY exchange interactions, crystalline electric field, and Kondo effect evolves from the bulk to nanowires of CeIn$_{3}$.  

These observations indicate that the dimensionality change in this compound led to the presence of magnetic frustration, opening a possibility to explore the role of dimensionality in the properties of strongly correlated materials.

\section{\label{sec:level2}Experimental methods}
\subsection{Synthesis method}

The CeIn$_{3}$ nanowires were grown by the Metallic Flux Nanonucleation (MFNN) method \cite{PIROTA202061,Patente}.  This method is a combination of the conventional metallic flux-growth technique \cite{Fisk} with the addition of a nanometric template for nanowire ingrowth, which provides simultaneous growth of nanowires and bulk crystals in the same batch, facilitating the comparative study of dimensionality effect on the properties of the compounds. Particularly, in this work, we used as template an Al$_{2}$O$_{3}$ membrane with a pore size of 155(25) nm fabricated by the hard anodization process described in detail in ref \cite{lee2006fast}. The template was added together with the metals in a ratio of 1 Ce:10 In inside a quartz tube, with the addition of quartz wool at the extremities of the tube to remove excess In. The sealed evacuated tube was placed in a furnace and heated up to $1100 ^{\circ}\mathrm{C}$ at a rate of $50^{\circ}\mathrm{C/h}$. After 8h at $1100 ^{\circ}\mathrm{C}$, the batch was cooled to $800^{\circ}\mathrm{C}$ at a rate of $1^{\circ}\mathrm{C/h}$, and finally to $650^{\circ}\mathrm{C}$ at $10^{\circ}\mathrm{C/h}$. The excess In flux was then spun in a centrifuge and the membrane with nanowires and the CeIn$_{3}$ crystals were mechanically removed from the quartz tube.

\subsection{Characterization}
CeIn$_{3}$ bulk and nanowires were analyzed with a commercial Quanta FEG 250 Field Emission Scanning Electron Microscope (FE-SEM) and submitted to elemental analysis using a commercial Energy Dispersive X-Ray Spectroscopy (EDS) microprobe model X-Max50. The selected area electron diffraction (SAED) patterns in the transmission electron microscope (TEM) were performed in a FEI Titan ETEM with image Cs corrector operating at 300 keV equipped with a Gatan K3-IS direct detection counting camera. For SAED analysis, a lamella was prepared from the CeIn$_{3}$ nanowires using a dual-beam focused ion beam/scanning electron microscope (FIB) from ThermoFisher Helios 600 Nanolab Ga$^+$ FIB/SEM - equipped with an OmniProbe micro-manipulator. Magnetization data were collected using a superconducting quantum interference device (SQUID) magnetometer MPMS-7T. Specific heat data (12 K to 0.3 K) were taken in a commercial Quantum Design PPMS using a $^3$He system.
NMR measurements were performed on a $^{3}$He cryostat equipped with a variable 17 T superconducting magnet at the National High Magnetic Field Laboratory (NHMFL) in Tallahassee-FL. A radio frequency coil was manufactured with a formvar insulating copper wire to perform low-$T$ NMR measurements at $40-50$ MHz. The same alumina template with the CeIn$_3$ nanowires used in the bulk measurements was mounted on a NMR probe equipped with a goniometer for fine alignment with the external field. The field-swept $^{115}$In NMR spectra ($I$ = 9/2; $\gamma/2\pi$ = 9.3295 MHz/T) were obtained by stepwise summing the Fourier transform of the spin-echo signal. The NMR data were collected using a MagRes2000 homodyne spectrometer.

\section{RESULTS AND DISCUSSION}
\subsection{Electron microscopy}

\begin{figure*}[ht!]
 \centering
 \includegraphics[width=\textwidth]{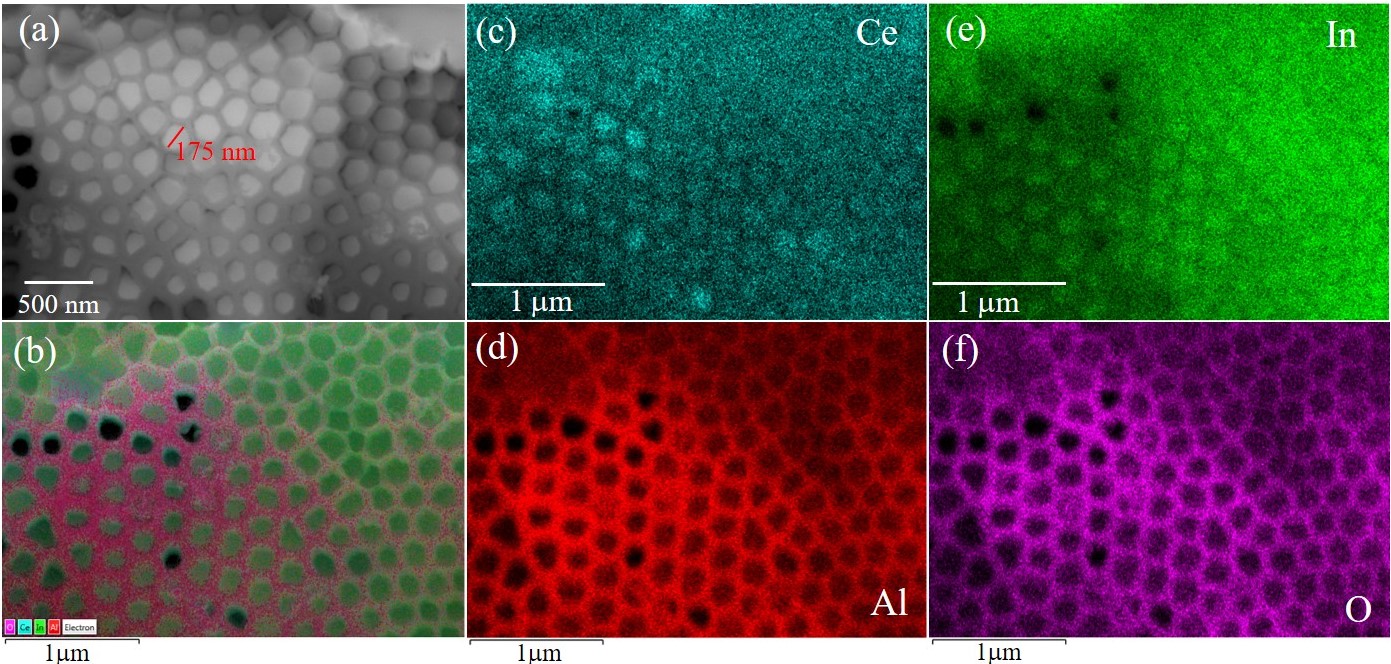}
 \caption{(a) Electron backscatter diffraction (EBSD) image showing the atomic number contrast of the CeIn$_{3}$ nanowires and the Al$_{2}$O$_{3}$ template;
\textbf{(b)} Image reconstructed from the EDS mapping of Figures (c-f).}
 \label{Fig1}
\end{figure*}

\begin{figure}[ht!]
\centering\includegraphics[width=1\linewidth]{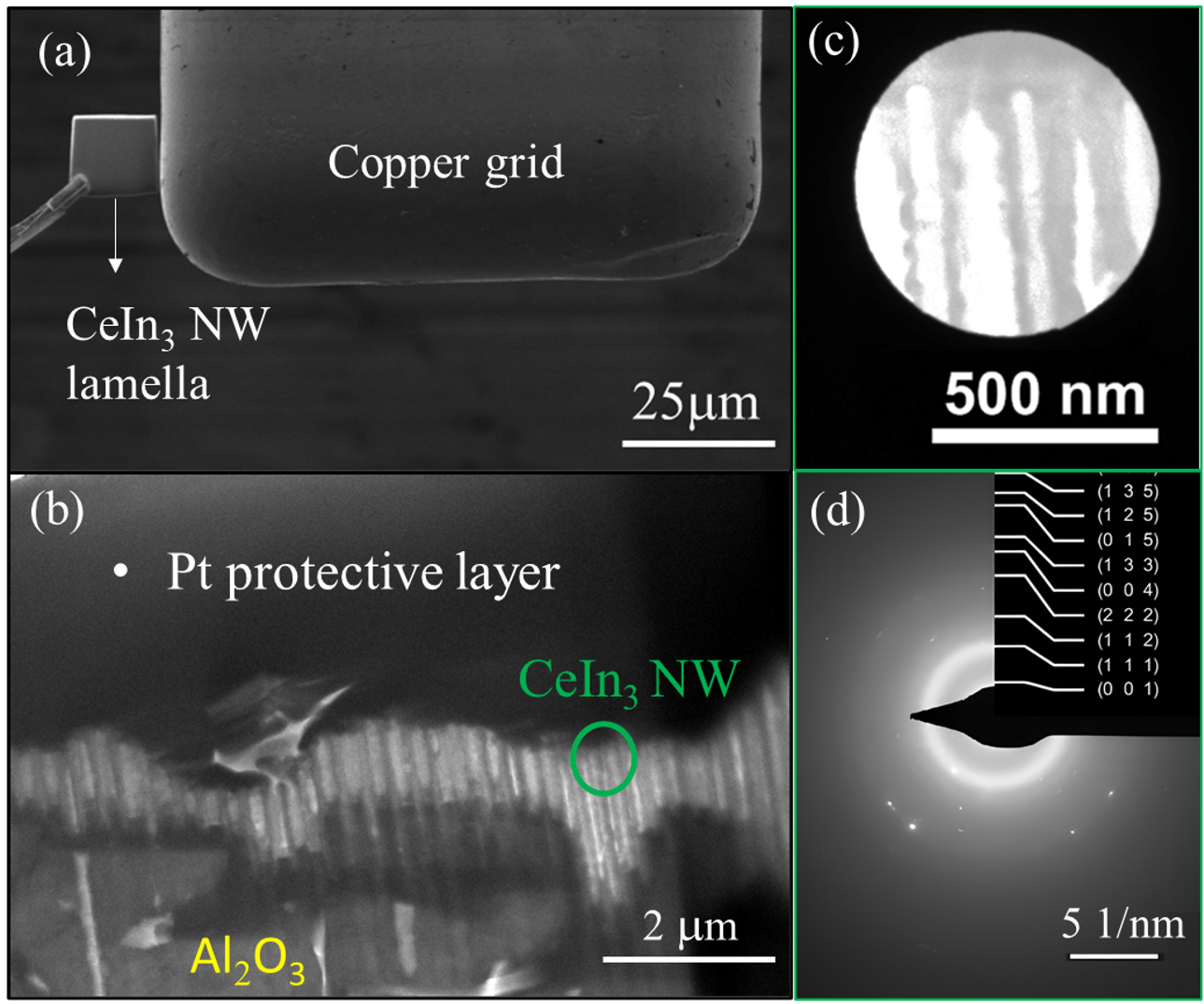}
\caption{(a) SEM image of a CeIn$_{3}$ lamella. (b) TEM image of the cross-section of the nanowires, showing the contrast of the Pt layer, the CeIn$_{3}$ nanowires, and the Al$_{2}$O$_{3}$ template. (c,d) SAED pattern of the nanowires. Inset highlights the region where the SAED was performed.}
\label{Fig2}
\end{figure}

Fig.~\ref{Fig1} displays the high resolution scanning electron microscope (FE-SEM) image and Energy Dispersive X-ray Spectrometry (EDS) mapping of a small area of the  Al$_{2}$O$_{3}$  template containing nanowires. Fig 1a shows an electron backscatter diffraction (EBSD) image that clearly reveals the atomic number contrast of the regions filled with CeIn$_{3}$ nanowires and the Al$_{2}$O$_{3}$ template. We observed the presence of CeIn$_{3}$ nanowires with an average diameter of $175(5)$ nm and a relatively homogeneous filling factor. The EDS characterization (Fig. 1c-f) confirms that the synthesized sample is composed only of Ce, In, Al and O. The semi-quantitative analysis of the atomic composition performed experimentally for Ce and In 1:3.1(1) and nominally 1:3 is in good agreement indicating the formation of CeIn$_{3}$. For SAED analysis, a lamella was prepared with a cross-sectional cut inside the CeIn$_{3}$ nanowires (Fig. 1a) using Ga$^+$ FIB, and placed on a copper (Cu) TEM grid, as illustrated in Fig.~\ref{Fig2}(a). To reduce beam-induced damage, the final thinning steps involved low kV (2 kV) thinning with a current of 27 pA. Fig (2b) shows a TEM image of the CeIn$_{3}$ lamella, displaying the cross section of the nanowires, which evidences a length distribution $\approx 1-2 \, \mu$m. 

We performed SAED on the region of Fig. 2(c) with at least 4 nanowires, to investigate the crystal structure and obtain the lattice parameters. The SAED pattern, Fig. 2(d), presents a polycrystalline characteristic, and the indexing reveals the presence of the cubic structure of CeIn$_{3}$ with space group Pm-3m. We obtained the lattice parameter \textit{a} = b = c = 4.78 \AA, in agreement with the literature \cite{Moshopoulou:ws5027}.

\subsection{Physical Characterization}

\begin{figure}[ht!]
\centering\includegraphics[width=1\linewidth]{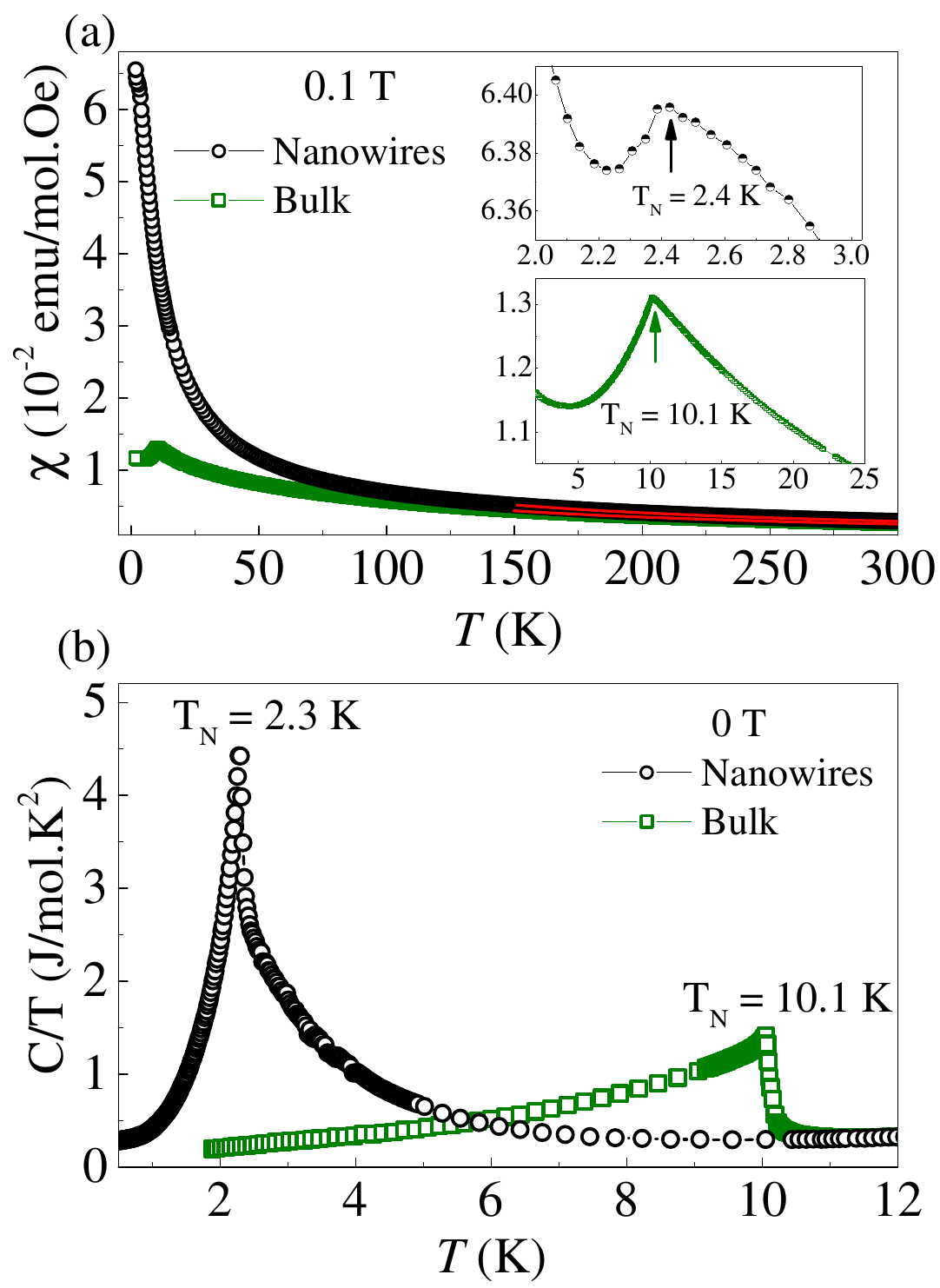}
\caption{Physical properties of bulk and nanowire CeIn$_{3}$ systems. {(a)} Temperature dependence of the magnetic susceptibility taken with applied field of $H=0.1$ T. {(b)} Specific heat divided by temperature as a function of temperature performed in 0 T. The inset shows the suppression of the N\'eel  temperature with reducing dimensionality.}
\label{Fig3}
\end{figure}

 Now we turn our attention to the role of dimensionality on the physical properties of both CeIn$_{3}$ systems. For the nanowire array characterization, we measured susceptibility $\chi$($T$) with the magnetic field applied parallel and perpendicular to the nanowire axis, but we present in Fig.~\ref{Fig3}(a) only the polycrystalline average of the  $\chi$($T$) as a function of temperature, for $H=0.1$ T. In these data, the paramagnetic contribution to the susceptibility from the Al$_{2}$O$_{3}$ template \cite{Carvalho_2022} has been subtracted from the total susceptibility, and the resulting difference is normalized by the molar mass of CeIn$_{3}$. We observed a drastic suppression of the antiferromagnetic ordering temperature from $T_{N}^{bulk} = 10.1$ K (inset green arrow) to $T_{N}^{nano} = 2.4$ K (inset black arrow). For both bulk and LD systems, $\chi$($T$) above 150 K can be fitted to a Curie-Weiss law plus a T-independent Pauli term, $\chi$($T$) = $\chi_{0}$ + $C$/($T - \theta_{CW})$ (solid lines). Table 1 displays
the fitting parameters. The effective moment ($p_{eff}$) for the Ce$^{3+}$ ions, extracted from the Curie-Weiss constant (C), is in very good agreement with $p_{eff}^{theory} = 2.54$ $\mu_{B}$ from the Hund's rule moment for free J = 5/2 cerium ions \cite{blundell2001magnetism}. This, along with the lattice constants obtained from SAED analyses, confirms the formation of CeIn$_{3}$ nanowires in the Al$_{2}$O$_{3}$ template. The negative sign of $\theta_{CW}$ indicates an antiferromagnetic interaction, consistent with an antiferromagnetic ground state of CeIn$_{3}$. 

Interestingly, the reduction of the high-$T$ $\theta_{CW}$ for the nanowires already indicates a clear suppression of the long range RKKY interaction in the CeIn$_{3}$ nanowires.  

Additionally, $|\theta_{CW}| > T_{N}$ for both systems, indicating the presence of magnetic frustration, as previously observed for GdIn$_{3}$ nanowires \cite{ROSA201414}. The ratio $\theta_{CW}/T_{N}$ is higher for the LD system, implying a more pronounced frustration in the nanowire than in the bulk system. However, the most significant finding from our data is the substantial reduction of $T_{N}$ by a factor of 4 in the nanowire system compared to the bulk system. This decrease is probably associated with a combined change in the contributions of the RKKY exchange interaction, the CEF, and the Kondo effect.

\begin{table}[h]
\small
  \caption{Curie-Weiss fitting parameters for both bulk and nanowire CeIn$_{3}$ systems with $\chi$($T$) = $\chi_{0}$ + C/($T - \theta_{CW})$.}
  \label{tbl:example1}
  \begin{tabular*}{0.48\textwidth}{@{\extracolsep{\fill}}lllll}
    \hline
    CeIn$_{3}$ & $\theta_{CW} (K)$ & $\chi_{0}$ (emu/mol.Oe) & p$_{eff}$ ($\mu_{B}$)  & $\frac{\theta_{CW}}{T_{N}} $ \\
    \hline
    Bulk & -43(1) & $2.0 (1) \times 10^{-4}$ & 2.54(1) & 4 \\
    Nanowire & -24(4) & $3.5  (6) \times 10^{-4}$ & 2.59(5) & 10 \\
    \hline
  \end{tabular*}
\end{table}

The suppression of the magnetic order at $T_{N}$ is further confirmed by specific heat measurements.
The curves of total specific heat \textit{C} divided by temperature as a function of temperature, \textit{C/T} $\times$ \textit{T}, for CeIn$_3$ nanowires and bulk, are displayed in Fig.~\ref{Fig3}(b). The sharp peaks in \textit{C/T} corresponding to the onset of the AFM order can be seen at 10.1 K for the bulk compound and at 2.3 K for the nanowire compound, in good agreement with the maximum observed in the magnetic susceptibility data (see Fig.~\ref{Fig3}(a)).

\begin{figure}[ht!]
\centering\includegraphics[width=1\linewidth]{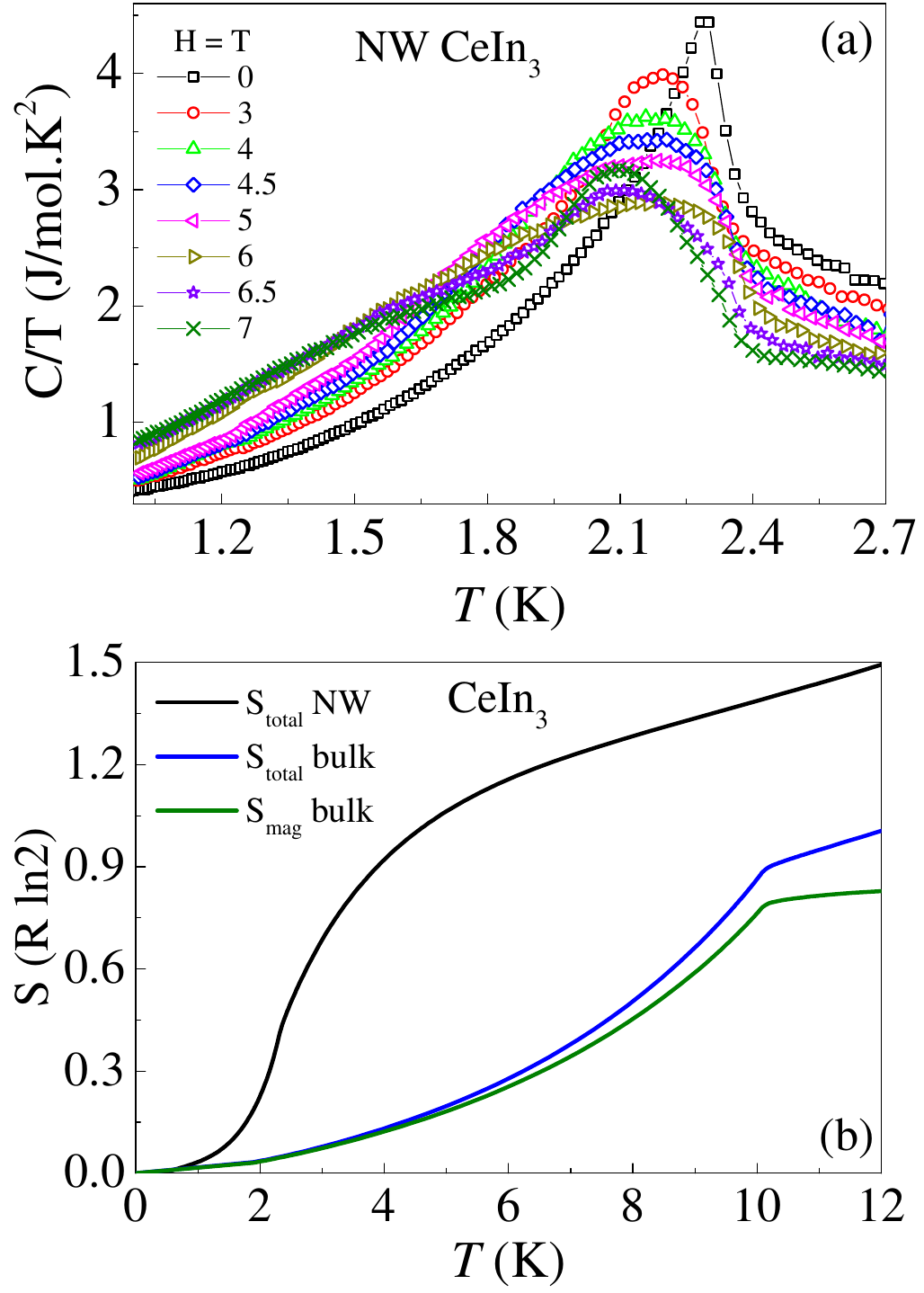}
\caption{(a) Total specific heat of CeIn$_{3}$ nanowires, plotted as $C/T \times T$, and (b) Total entropy for the NW and bulk system of the specific heat data displayed in Fig. 3(b), and magnetic entropy (after subtracting the lattice contribution) only for CeIn$_{3}$ bulk.}
\label{Fig4}
\end{figure}

In Fig.~\ref{Fig4} we show \textit{C/T} for different values of applied magnetic field and the estimated magnetic entropy $S$ as a function of $T$ for the CeIn$_3$ nanowires. It is interesting to note that the magnetic transition in Fig.~\ref{Fig4}(a) broadens and changes its shape qualitatively. Further, the onset of the AFM order shifts towards lower temperatures upon increasing the magnetic field. This broadening is possibly associated with the presence of an inhomogeneous shift of $T_{N}$ lower than $T$ due to disorder and/or diameter/internal field distribution in the arrangement of nanowires. In addition, a double peak feature can be observed for \textit{H} ${\geq}$ 6 T, which may also be a manifestation of disorder or, most likely, can be associated with a magnetic field-induced phase transition, similar to what was reported for the 115 compounds in \cite{duque2011field}.
The corresponding total entropy $[ S=\int (C/T)\, dT]$ is obtained by integrating the total contribution to the specific heat up to 12 K, and shown as a function of T in Fig. 4(b).
We also show the magnetic entropy $S_{mag}$ for CeIn$_{3}$ single crystals obtained after subtraction of the non-magnetic contribution of LaIn$_3$ from the C$_p$ data. The $S_{mag}$ (12 K) is $\approx83\%$ of $R \ln 2$ for CeIn$_{3}$ suggesting a slightly compensated doublet CEF ground-state for the Ce$^{3+}$ ion, in agreement with previous reports \cite{pagliuso2006,lawrence1980,knafo2003}.

On the other hand, for CeIn$_{3}$ NW, the entropy recovered at $T_{N}$ is only roughly 0.5 of $R \ln 2$, which indicates a higher Kondo compensation of the CEF doublet ground state for the NW. Furthermore, the evolution of entropy shows that roughly 1.5 $R \ln 2$ at 12 K, which indicates that the excited CEF $\Gamma_{8}$ quartet state is much lower in energy for the nanowires compared to bulk CeIn$_{3}$\cite{pagliuso2006,lawrence1980,knafo2003}. The reduction of the energy of the CEF excited states has been claimed as one of the possible mechanisms for the reduction of $T_{N}$ for the tetragonal AFM analogs Ce$_{m}$RhIn$_{3m+2}$ ($m$ = 1,2) when compared to CeIn$_{3}$. 
This result should be taken with some care, since the nonmagnetic reference LaIn$_{3}$ for the NW system is not available to estimate phononic contributions in this system. 

\subsection{Nuclear Magnetic Resonance}

\begin{figure*}[ht!]
\centering\includegraphics[width=0.9\textwidth]{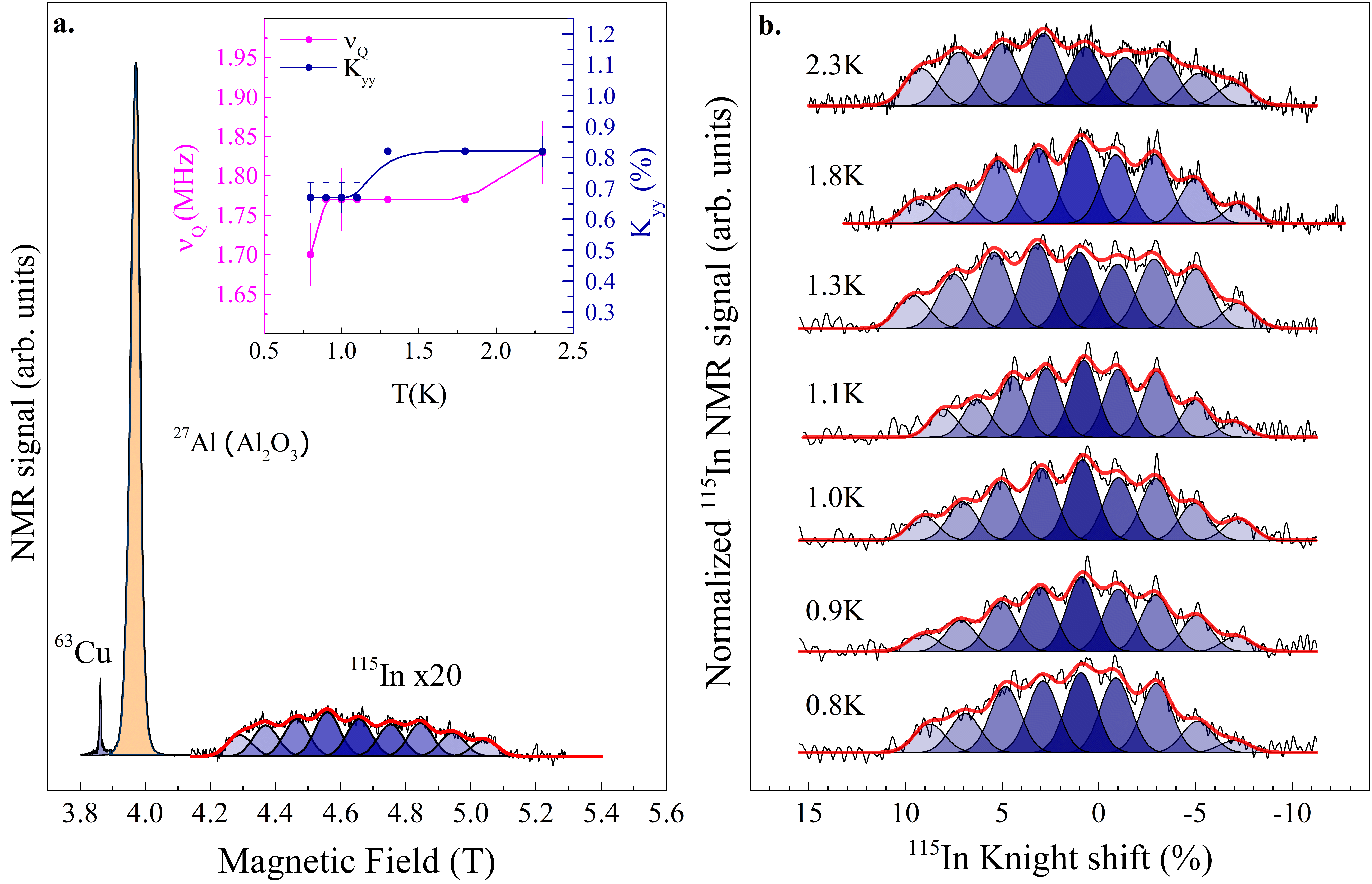}
\caption{(a) FFT field swept NMR spectrum at 2.3 K showing the $^{63}$Cu NMR signal from the coil, the $^{27}$Al NMR signal from the alumina template and the weak $^{115}$In NMR signal from the CeIn$_{3}$ nanowires (multiplied 20 times). The inset highlights the Knight Shift $^{115}$K$_{yy}$ and the quadrupole resonance frequency $\nu_{Q}$ as a function of temperature; (b) Low-$T$ normalized $^{115}$In NMR spectra as a function of temperature. 
}
\label{Fig5}
\end{figure*}

\begin{figure*}[ht!]
\centering\includegraphics[width=0.9\textwidth]{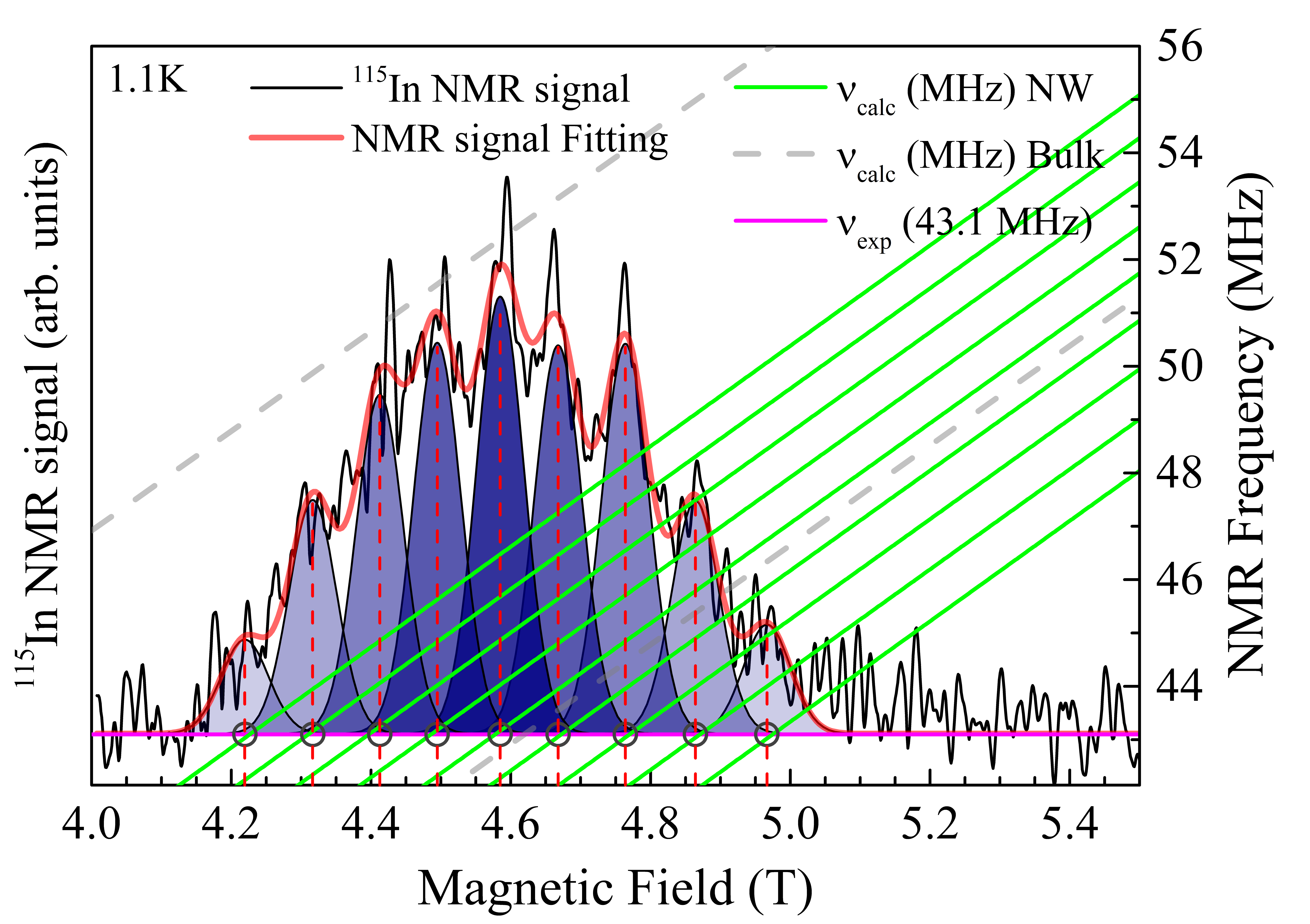}
\caption{$^{115}$In NMR signal and fitting. The solid lines indicate the calculated frequencies as a function of magnetic field for each resonance transition of the $I$ = 9/2 nuclear spin system calculated after full diagonalization of the nuclear Hamiltonian. The gray (green) solid lines are calculated considering a $\nu_Q$ = 9.6 MHz ($\nu_Q$ = 1.77(2) MHz) for the bulk (nanowires); 
}
\label{Fig6}
\end{figure*}

The full NMR spectrum measured at 2.3 K is shown in Fig.~\ref{Fig5}.a. It was possible to index 3 main contributions: the $^{63}$Cu signal from the coil used in the measurement, the strong $^{27}$Al signal referred to the Al$_{2}$O$_{3}$ template used in the growth method and the weak $^{115}$In signal referred to the CeIn$_{3}$ nanowires. Its intensity was increased by twenty times, so it would be noticeable in the image.

One can notice a great difference between the intensity of the $^{27}$Al NMR signal and the $^{115}$In NMR signal, which is mainly due to the amount of active nuclei in each specimen. Most of the NMR signal observed is from the alumina template because of its larger mass, as expected. In the chosen frequency/field range, the $^{27}$Al and $^{63}$Cu NMR signals did not overlap and were used as references for the $^{115}$In NMR signal of our CeIn$_{3}$ nanowires. 

Figure \ref{Fig5}.b presents the $^{115}$In NMR spectra at various temperatures. From the best fits of the spectra with a convolution of 9 Gaussian lines, we extracted the values of Knight shift, linewidths, and $\nu_Q$ as a function of temperature at the low-$T$ regime. The $K$ and $\nu_Q$ temperature dependencies are shown in the inset of Fig\ref{Fig5}.a.

Basically, a noticeable change can be seen in both parameters around $T\approx 1.8$ K at roughly 4.65 T, in agreement with the antiferromagnetic order observed by specific heat measurements under a magnetic field for our CeIn$_3$ nanowires. 

Fig.~\ref{Fig6} presents the $^{115}$In NMR spectra obtained by sweeping up $H\parallel$ Al$_{2}$O$_{3}$ template at constant frequency $\nu$ = 43.1 MHz. The spectra present the typical features of a nuclear spin $I$ = 9/2 with Zeeman and quadrupole couplings.

At low temperatures, the spectra show a broad central line with a full width at half maximum (FWHM) of {839} kHz (8.9 kOe) and four others broad satellite lines from each side split by the quadrupole interaction of the In nuclei with the local electric field gradient (EFG). The broad lines are due to the strain introduced by disorder, the different number of Ce neighbors to In sites, and the broadening is due to the distribution of both hyperfine fields and EFG.

The quadrupole splitting is proportional to the EFG at the In site, $V_{zz}$, through the relation: $\nu_Q = eQV_{zz}/2\sqrt{1+\eta^2/3}$, where Q is the nuclear quadrupole moment of $^{115}$In and $\eta$ is the asymmetry parameter of the EFG ($\eta$ = 0 for a tetragonal site symmetry). $V_{zz}$ arises from the hybridization between the In-5$d$ and the Ce-4$f$ orbitals resulting in the quadrupole frequency, $\nu_Q$ = 1.77(2) MHz. Strikingly, this is roughly 5 times smaller than $\nu_Q$ = 9.6 MHz observed in the bulk CeIn$_3$ single crystal \cite{kohori1999115in}. This suppression in the quadrupole frequency suggests a major change in the EFG, and thus in the crystallographic parameters at low temperatures. This result is consistent with the changes in the CEF effects and the consequent changes in the Ce 4$f$ CEF scheme of energy levels also suggested by the heat capacity data.  

In order to fully characterize the $^{115}$In spectra, we have calculated the frequency as a function of applied magnetic field utilizing a model of exact diagonalization of the $\textbf{Zeeman + Quadrupole}$ Hamiltonian using the software $\emph{MagRes2000}$ 8.9. In Fig. \ref{Fig6} one can see the results of these simulations. We adjusted the Knight shift $K_{yy}$ and the $\nu_Q$, so that we had a match between the perceived peak positions of the $^{115}$In NMR spectra and the center of the simulated curves projected at the constant measured frequency (magenta line at 43.1 MHz). The agreement between our results can be seen through the points of intersection between the transversal green lines and the horizontal magenta one. These lines represent the energy of the transition between the In nuclei spin energy levels and the frequency at which the NMR experiment was carried out ($43.1$ MHz), respectively. Moreover, we also show in Fig.\ref{Fig6} the In nuclei spin energy transitions for the CeIn$_3$ bulk sample through the gray dashed transversal lines, for comparison. This evidences the drastic suppression of $\nu_Q$ from 9.6 MHz (bulk) to 1.77(2) MHz (nanowires).

Furthermore, if one compares the suppression of $T_N$ with the reduction in $\nu_Q$, from bulk to nanowires, one verifies a similar factor of 4-5 times. In other words, $T_N^{bulk}/T_N^{nanowire} \approx 4.2$ is comparable to $\nu_q^{bulk}/\nu_q^{nanowire} \approx 5.5$, suggesting a possible correlation between both parameters.

It is worth noting that to obtain these calculated curves, we had to assume a $90\,^\circ$ Euler angle ($\theta_H$) between the applied magnetic field and the crystallographic $c$-axis, meaning that our CeIn$_3$ nanowires are most likely polycrystalline, as also suggested by the SAED data. Therefore, this is a hint that the growth of the nanowires inside the alumina template was not quite aligned along the uniaxial direction of the nanopores.

The NMR results reported here clearly indicate a significant influence of dimensionality on the magnetic properties of CeIn$_3$ and provide microscopic evidence for dimensionality-induced magnetic order frustration. Low-$T$ NMR supplies proof for polycrystalline nanonucleation of the CeIn$_3$ nanowires and explains the suppression of $T_N$ as a frustration of the magnetic order induced by reducing dimensionality.

As our data gives evidence for the reduction of the RKKY interaction, changes in the CEF and Kondo effects from the bulk to CeIn$_3$ nanowires. Presumably, all these effects are contributing to the frustration of the magnetic order in the CeIn$_3$ nanowires.

\FloatBarrier

\section{CONCLUSIONS}

In this work, we synthesized the intermetallic CeIn$_{3}$ nanowires by the Metallic-Flux Nanonucleation method and studied the effects of dimensionality via macroscopic and microscopic measurements. The SAED pattern indexing confirmed that the nanowires show the cubic phase of CeIn$_{3}$. Magnetic susceptibility, specific heat, and NMR data showed a drastic suppression of the antiferromagnetic transition from the bulk ($T^{3D}_{N} = 10.1$ K) to the nanowire system ($T^{LD}_{N} = 2.3$ K), which may be associated with a change in the RKKY magnetic exchange interaction, crystal electric field, and the Kondo effect. Our findings suggest that the dimensionality change in this compound led to the presence of magnetic frustration and motivates the use of the Metallic-Flux Nanonucleation method to study dimensionality effects more broadly in quantum materials.

\section{Acknowledgments}

This work was financially supported by the Brazilian funding agencies Fundação de
Amparo à Pesquisa do Estado de São Paulo, FAPESP (Grants
No 2022/16823-5 2020/10580-8,  2017/10581-1) and CNPq. We are grateful for the electronic microscopy support
from the LNNano – Brazilian Nanotechnology National Laboratory (LNNano/CNPEM/MCTIC) and Center for Integrated Nanotechnology (CINT). The authors thank Winson Kuo, for the experimental support in the Focused Ion Beam, Priscila F. S. Rosa, and Joe Thompson for important contributions to this work.

\nocite{*}

\bibliography{CeIn3}

\begin{thebibliography}{38}%
\makeatletter
\providecommand \@ifxundefined [1]{%
 \@ifx{#1\undefined}
}%
\providecommand \@ifnum [1]{%
 \ifnum #1\expandafter \@firstoftwo
 \else \expandafter \@secondoftwo
 \fi
}%
\providecommand \@ifx [1]{%
 \ifx #1\expandafter \@firstoftwo
 \else \expandafter \@secondoftwo
 \fi
}%
\providecommand \natexlab [1]{#1}%
\providecommand \enquote  [1]{``#1''}%
\providecommand \bibnamefont  [1]{#1}%
\providecommand \bibfnamefont [1]{#1}%
\providecommand \citenamefont [1]{#1}%
\providecommand \href@noop [0]{\@secondoftwo}%
\providecommand \href [0]{\begingroup \@sanitize@url \@href}%
\providecommand \@href[1]{\@@startlink{#1}\@@href}%
\providecommand \@@href[1]{\endgroup#1\@@endlink}%
\providecommand \@sanitize@url [0]{\catcode `\\12\catcode `\$12\catcode `\&12\catcode `\#12\catcode `\^12\catcode `\_12\catcode `\%12\relax}%
\providecommand \@@startlink[1]{}%
\providecommand \@@endlink[0]{}%
\providecommand \url  [0]{\begingroup\@sanitize@url \@url }%
\providecommand \@url [1]{\endgroup\@href {#1}{\urlprefix }}%
\providecommand \urlprefix  [0]{URL }%
\providecommand \Eprint [0]{\href }%
\providecommand \doibase [0]{https://doi.org/}%
\providecommand \selectlanguage [0]{\@gobble}%
\providecommand \bibinfo  [0]{\@secondoftwo}%
\providecommand \bibfield  [0]{\@secondoftwo}%
\providecommand \translation [1]{[#1]}%
\providecommand \BibitemOpen [0]{}%
\providecommand \bibitemStop [0]{}%
\providecommand \bibitemNoStop [0]{.\EOS\space}%
\providecommand \EOS [0]{\spacefactor3000\relax}%
\providecommand \BibitemShut  [1]{\csname bibitem#1\endcsname}%
\let\auto@bib@innerbib\@empty
\bibitem [{\citenamefont {Ashcroft}\ and\ \citenamefont {Mermin}(1976)}]{ashcroft1976solid}%
  \BibitemOpen
  \bibfield  {author} {\bibinfo {author} {\bibfnamefont {N.~W.}\ \bibnamefont {Ashcroft}}\ and\ \bibinfo {author} {\bibfnamefont {N.~D.}\ \bibnamefont {Mermin}},\ }\href@noop {} {\emph {\bibinfo {title} {Solid State Physics}}}\ (\bibinfo  {publisher} {Saunders College},\ \bibinfo {address} {Philadelphia, PA},\ \bibinfo {year} {1976})\BibitemShut {NoStop}%
\bibitem [{\citenamefont {Fetter}\ and\ \citenamefont {Walecka}(2003)}]{fetter2003quantum}%
  \BibitemOpen
  \bibfield  {author} {\bibinfo {author} {\bibfnamefont {A.}~\bibnamefont {Fetter}}\ and\ \bibinfo {author} {\bibfnamefont {J.}~\bibnamefont {Walecka}},\ }\href {https://books.google.com.br/books?id=0wekf1s83b0C} {\emph {\bibinfo {title} {Quantum Theory of Many-particle Systems}}},\ Dover Books on Physics\ (\bibinfo  {publisher} {Dover Publications},\ \bibinfo {year} {2003})\BibitemShut {NoStop}%
\bibitem [{\citenamefont {Berry}\ \emph {et~al.}(2010)\citenamefont {Berry}, \citenamefont {Bittar}, \citenamefont {Capan}, \citenamefont {Pagliuso},\ and\ \citenamefont {Fisk}}]{berry2010magnetic}%
  \BibitemOpen
  \bibfield  {author} {\bibinfo {author} {\bibfnamefont {N.}~\bibnamefont {Berry}}, \bibinfo {author} {\bibfnamefont {E.}~\bibnamefont {Bittar}}, \bibinfo {author} {\bibfnamefont {C.}~\bibnamefont {Capan}}, \bibinfo {author} {\bibfnamefont {P.}~\bibnamefont {Pagliuso}},\ and\ \bibinfo {author} {\bibfnamefont {Z.}~\bibnamefont {Fisk}},\ }\bibfield  {title} {\bibinfo {title} {Magnetic, thermal, and transport properties of cd-doped cein$_3$},\ }\href@noop {} {\bibfield  {journal} {\bibinfo  {journal} {Physical Review B-Condensed Matter and Materials Physics}\ }\textbf {\bibinfo {volume} {81}},\ \bibinfo {pages} {174413} (\bibinfo {year} {2010})}\BibitemShut {NoStop}%
\bibitem [{\citenamefont {Ebihara}\ \emph {et~al.}(2004)\citenamefont {Ebihara}, \citenamefont {Harrison}, \citenamefont {Jaime}, \citenamefont {Uji},\ and\ \citenamefont {Lashley}}]{ebihara2004emergent}%
  \BibitemOpen
  \bibfield  {author} {\bibinfo {author} {\bibfnamefont {T.}~\bibnamefont {Ebihara}}, \bibinfo {author} {\bibfnamefont {N.}~\bibnamefont {Harrison}}, \bibinfo {author} {\bibfnamefont {M.}~\bibnamefont {Jaime}}, \bibinfo {author} {\bibfnamefont {S.}~\bibnamefont {Uji}},\ and\ \bibinfo {author} {\bibfnamefont {J.}~\bibnamefont {Lashley}},\ }\bibfield  {title} {\bibinfo {title} {Emergent fluctuation hot spots on the fermi surface of c e i n 3 in strong magnetic fields},\ }\href@noop {} {\bibfield  {journal} {\bibinfo  {journal} {Physical review letters}\ }\textbf {\bibinfo {volume} {93}},\ \bibinfo {pages} {246401} (\bibinfo {year} {2004})}\BibitemShut {NoStop}%
\bibitem [{\citenamefont {Moll}\ \emph {et~al.}(2017{\natexlab{a}})\citenamefont {Moll}, \citenamefont {Helm}, \citenamefont {Zhang}, \citenamefont {Batista}, \citenamefont {Harrison}, \citenamefont {McDonald}, \citenamefont {Winter}, \citenamefont {Ramshaw}, \citenamefont {Chan}, \citenamefont {Balakirev} \emph {et~al.}}]{moll2017emergent}%
  \BibitemOpen
  \bibfield  {author} {\bibinfo {author} {\bibfnamefont {P.~J.}\ \bibnamefont {Moll}}, \bibinfo {author} {\bibfnamefont {T.}~\bibnamefont {Helm}}, \bibinfo {author} {\bibfnamefont {S.-S.}\ \bibnamefont {Zhang}}, \bibinfo {author} {\bibfnamefont {C.~D.}\ \bibnamefont {Batista}}, \bibinfo {author} {\bibfnamefont {N.}~\bibnamefont {Harrison}}, \bibinfo {author} {\bibfnamefont {R.~D.}\ \bibnamefont {McDonald}}, \bibinfo {author} {\bibfnamefont {L.~E.}\ \bibnamefont {Winter}}, \bibinfo {author} {\bibfnamefont {B.}~\bibnamefont {Ramshaw}}, \bibinfo {author} {\bibfnamefont {M.~K.}\ \bibnamefont {Chan}}, \bibinfo {author} {\bibfnamefont {F.~F.}\ \bibnamefont {Balakirev}}, \emph {et~al.},\ }\bibfield  {title} {\bibinfo {title} {Emergent magnetic anisotropy in the cubic heavy-fermion metal cein3},\ }\href@noop {} {\bibfield  {journal} {\bibinfo  {journal} {npj Quantum Materials}\ }\textbf {\bibinfo {volume} {2}},\ \bibinfo {pages} {46} (\bibinfo {year} {2017}{\natexlab{a}})}\BibitemShut {NoStop}%
\bibitem [{\citenamefont {Mathur}\ \emph {et~al.}(1998)\citenamefont {Mathur}, \citenamefont {Grosche}, \citenamefont {Julian}, \citenamefont {Walker}, \citenamefont {Freye}, \citenamefont {Haselwimmer},\ and\ \citenamefont {Lonzarich}}]{mathur1998magnetically}%
  \BibitemOpen
  \bibfield  {author} {\bibinfo {author} {\bibfnamefont {N.}~\bibnamefont {Mathur}}, \bibinfo {author} {\bibfnamefont {F.}~\bibnamefont {Grosche}}, \bibinfo {author} {\bibfnamefont {S.}~\bibnamefont {Julian}}, \bibinfo {author} {\bibfnamefont {I.}~\bibnamefont {Walker}}, \bibinfo {author} {\bibfnamefont {D.}~\bibnamefont {Freye}}, \bibinfo {author} {\bibfnamefont {R.}~\bibnamefont {Haselwimmer}},\ and\ \bibinfo {author} {\bibfnamefont {G.}~\bibnamefont {Lonzarich}},\ }\bibfield  {title} {\bibinfo {title} {Magnetically mediated superconductivity in heavy fermion compounds},\ }\href@noop {} {\bibfield  {journal} {\bibinfo  {journal} {Nature}\ }\textbf {\bibinfo {volume} {394}},\ \bibinfo {pages} {39} (\bibinfo {year} {1998})}\BibitemShut {NoStop}%
\bibitem [{\citenamefont {Grosche}\ \emph {et~al.}(2001)\citenamefont {Grosche}, \citenamefont {Walker}, \citenamefont {Julian}, \citenamefont {Mathur}, \citenamefont {Freye}, \citenamefont {Steiner},\ and\ \citenamefont {Lonzarich}}]{grosche2001superconductivity}%
  \BibitemOpen
  \bibfield  {author} {\bibinfo {author} {\bibfnamefont {F.}~\bibnamefont {Grosche}}, \bibinfo {author} {\bibfnamefont {I.}~\bibnamefont {Walker}}, \bibinfo {author} {\bibfnamefont {S.}~\bibnamefont {Julian}}, \bibinfo {author} {\bibfnamefont {N.}~\bibnamefont {Mathur}}, \bibinfo {author} {\bibfnamefont {D.}~\bibnamefont {Freye}}, \bibinfo {author} {\bibfnamefont {M.}~\bibnamefont {Steiner}},\ and\ \bibinfo {author} {\bibfnamefont {G.}~\bibnamefont {Lonzarich}},\ }\bibfield  {title} {\bibinfo {title} {Superconductivity on the threshold ofmagnetism in cepd2si2 and cein3},\ }\href@noop {} {\bibfield  {journal} {\bibinfo  {journal} {Journal of Physics: Condensed Matter}\ }\textbf {\bibinfo {volume} {13}},\ \bibinfo {pages} {2845} (\bibinfo {year} {2001})}\BibitemShut {NoStop}%
\bibitem [{\citenamefont {Shishido}\ \emph {et~al.}(2010)\citenamefont {Shishido}, \citenamefont {Shibauchi}, \citenamefont {Yasu}, \citenamefont {Kato}, \citenamefont {Kontani}, \citenamefont {Terashima},\ and\ \citenamefont {Matsuda}}]{shishido2010tuning}%
  \BibitemOpen
  \bibfield  {author} {\bibinfo {author} {\bibfnamefont {H.}~\bibnamefont {Shishido}}, \bibinfo {author} {\bibfnamefont {T.}~\bibnamefont {Shibauchi}}, \bibinfo {author} {\bibfnamefont {K.}~\bibnamefont {Yasu}}, \bibinfo {author} {\bibfnamefont {T.}~\bibnamefont {Kato}}, \bibinfo {author} {\bibfnamefont {H.}~\bibnamefont {Kontani}}, \bibinfo {author} {\bibfnamefont {T.}~\bibnamefont {Terashima}},\ and\ \bibinfo {author} {\bibfnamefont {Y.}~\bibnamefont {Matsuda}},\ }\bibfield  {title} {\bibinfo {title} {Tuning the dimensionality of the heavy fermion compound cein3},\ }\href@noop {} {\bibfield  {journal} {\bibinfo  {journal} {Science}\ }\textbf {\bibinfo {volume} {327}},\ \bibinfo {pages} {980} (\bibinfo {year} {2010})}\BibitemShut {NoStop}%
\bibitem [{\citenamefont {Paschen}\ and\ \citenamefont {Si}(2021)}]{paschen2021quantum}%
  \BibitemOpen
  \bibfield  {author} {\bibinfo {author} {\bibfnamefont {S.}~\bibnamefont {Paschen}}\ and\ \bibinfo {author} {\bibfnamefont {Q.}~\bibnamefont {Si}},\ }\bibfield  {title} {\bibinfo {title} {Quantum phases driven by strong correlations},\ }\href@noop {} {\bibfield  {journal} {\bibinfo  {journal} {Nature Reviews Physics}\ }\textbf {\bibinfo {volume} {3}},\ \bibinfo {pages} {9} (\bibinfo {year} {2021})}\BibitemShut {NoStop}%
\bibitem [{\citenamefont {{Joe D. Thompson and Zachary Fisk}}(2012)}]{d2011progress}%
  \BibitemOpen
  \bibfield  {author} {\bibinfo {author} {\bibnamefont {{Joe D. Thompson and Zachary Fisk}}},\ }\bibfield  {title} {\bibinfo {title} {{Progress in Heavy-Fermion Superconductivity: Ce115 and Related Materials}},\ }\href@noop {} {\bibfield  {journal} {\bibinfo  {journal} {Journal of the Physical Society of Japan}\ }\textbf {\bibinfo {volume} {81}},\ \bibinfo {pages} {011002} (\bibinfo {year} {2012})}\BibitemShut {NoStop}%
\bibitem [{\citenamefont {{C. Petrovic, R. Movshovich, M. Jaime, P. G. Pagliuso, M. F. Hundley, J. L. Sarrao, Z. Fisk and J. D. Thompson}}(2001)}]{petrovic1}%
  \BibitemOpen
  \bibfield  {author} {\bibinfo {author} {\bibnamefont {{C. Petrovic, R. Movshovich, M. Jaime, P. G. Pagliuso, M. F. Hundley, J. L. Sarrao, Z. Fisk and J. D. Thompson}}},\ }\bibfield  {title} {\bibinfo {title} {{A new heavy-fermion superconductor CeIrIn$_5$: A relative of the cuprates?}},\ }\href@noop {} {\bibfield  {journal} {\bibinfo  {journal} {Europhysics Letters}\ }\textbf {\bibinfo {volume} {53}},\ \bibinfo {pages} {354} (\bibinfo {year} {2001})}\BibitemShut {NoStop}%
\bibitem [{\citenamefont {{C Petrovic, P G Pagliuso, M F Hundley, R Movshovich, J L Sarrao, J D Thompson, Z Fisk and P Monthoux}}(2001)}]{petrovic2}%
  \BibitemOpen
  \bibfield  {author} {\bibinfo {author} {\bibnamefont {{C Petrovic, P G Pagliuso, M F Hundley, R Movshovich, J L Sarrao, J D Thompson, Z Fisk and P Monthoux}}},\ }\bibfield  {title} {\bibinfo {title} {{Heavy-fermion superconductivity in CeCoIn$_5$ at 2.3 K}},\ }\href@noop {} {\bibfield  {journal} {\bibinfo  {journal} {Journal of Physics: Condensed Matter}\ }\textbf {\bibinfo {volume} {13}},\ \bibinfo {pages} {L337} (\bibinfo {year} {2001})}\BibitemShut {NoStop}%
\bibitem [{\citenamefont {Buschow}\ \emph {et~al.}(1969)\citenamefont {Buschow}, \citenamefont {De~Wijn},\ and\ \citenamefont {Van~Diepen}}]{buschow1969}%
  \BibitemOpen
  \bibfield  {author} {\bibinfo {author} {\bibfnamefont {K.}~\bibnamefont {Buschow}}, \bibinfo {author} {\bibfnamefont {H.}~\bibnamefont {De~Wijn}},\ and\ \bibinfo {author} {\bibfnamefont {A.}~\bibnamefont {Van~Diepen}},\ }\bibfield  {title} {\bibinfo {title} {Magnetic susceptibilities of rare-earth--indium compounds: Rin3},\ }\href@noop {} {\bibfield  {journal} {\bibinfo  {journal} {The Journal of Chemical Physics}\ }\textbf {\bibinfo {volume} {50}},\ \bibinfo {pages} {137} (\bibinfo {year} {1969})}\BibitemShut {NoStop}%
\bibitem [{\citenamefont {Amara}\ \emph {et~al.}(1994)\citenamefont {Amara}, \citenamefont {Galera}, \citenamefont {Morin}, \citenamefont {Veres},\ and\ \citenamefont {Burlet}}]{amara1994}%
  \BibitemOpen
  \bibfield  {author} {\bibinfo {author} {\bibfnamefont {M.}~\bibnamefont {Amara}}, \bibinfo {author} {\bibfnamefont {R.}~\bibnamefont {Galera}}, \bibinfo {author} {\bibfnamefont {P.}~\bibnamefont {Morin}}, \bibinfo {author} {\bibfnamefont {T.}~\bibnamefont {Veres}},\ and\ \bibinfo {author} {\bibfnamefont {P.}~\bibnamefont {Burlet}},\ }\bibfield  {title} {\bibinfo {title} {Incommensurate antiferromagnetic structures in ndin3},\ }\href@noop {} {\bibfield  {journal} {\bibinfo  {journal} {Journal of magnetism and magnetic materials}\ }\textbf {\bibinfo {volume} {130}},\ \bibinfo {pages} {127} (\bibinfo {year} {1994})}\BibitemShut {NoStop}%
\bibitem [{\citenamefont {Amara}\ \emph {et~al.}(1995)\citenamefont {Amara}, \citenamefont {Gal{\'e}ra}, \citenamefont {Morin}, \citenamefont {Voiron},\ and\ \citenamefont {Burlet}}]{amara1995}%
  \BibitemOpen
  \bibfield  {author} {\bibinfo {author} {\bibfnamefont {M.}~\bibnamefont {Amara}}, \bibinfo {author} {\bibfnamefont {R.}~\bibnamefont {Gal{\'e}ra}}, \bibinfo {author} {\bibfnamefont {P.}~\bibnamefont {Morin}}, \bibinfo {author} {\bibfnamefont {J.}~\bibnamefont {Voiron}},\ and\ \bibinfo {author} {\bibfnamefont {P.}~\bibnamefont {Burlet}},\ }\bibfield  {title} {\bibinfo {title} {Magnetic phase diagram in ndin3},\ }\href@noop {} {\bibfield  {journal} {\bibinfo  {journal} {Journal of magnetism and magnetic materials}\ }\textbf {\bibinfo {volume} {140}},\ \bibinfo {pages} {1157} (\bibinfo {year} {1995})}\BibitemShut {NoStop}%
\bibitem [{\citenamefont {{P. G. Pagliuso, D. J. Garcia, E. Miranda, E. Granado, R. Lora Serrano, C. Giles, J. G. S. Duque, R. R. Urbano, C. Rettori, J. D. Thompson, M. F. Hundley and J. L. Sarrao}}(2006)}]{pagliuso2006}%
  \BibitemOpen
  \bibfield  {author} {\bibinfo {author} {\bibnamefont {{P. G. Pagliuso, D. J. Garcia, E. Miranda, E. Granado, R. Lora Serrano, C. Giles, J. G. S. Duque, R. R. Urbano, C. Rettori, J. D. Thompson, M. F. Hundley and J. L. Sarrao}}},\ }\bibfield  {title} {\bibinfo {title} {{Evolution of the magnetic properties and magnetic structures along the R$_m$MIn$\textsubscript{3m+2}$ (R = Ce, Nd, Gd, Tb; M = Rh, Ir; and m = 1, 2) series of intermetallic compounds}},\ }\href@noop {} {\bibfield  {journal} {\bibinfo  {journal} {Journal of Applied Physics}\ }\textbf {\bibinfo {volume} {99}},\ \bibinfo {pages} {08P703} (\bibinfo {year} {2006})}\BibitemShut {NoStop}%
\bibitem [{\citenamefont {Moll}\ \emph {et~al.}(2017{\natexlab{b}})\citenamefont {Moll}, \citenamefont {Helm}, \citenamefont {Zhang}, \citenamefont {Batista}, \citenamefont {Harrison}, \citenamefont {McDonald}, \citenamefont {Winter}, \citenamefont {Ramshaw}, \citenamefont {Chan}, \citenamefont {Balakirev} \emph {et~al.}}]{moll2017}%
  \BibitemOpen
  \bibfield  {author} {\bibinfo {author} {\bibfnamefont {P.~J.}\ \bibnamefont {Moll}}, \bibinfo {author} {\bibfnamefont {T.}~\bibnamefont {Helm}}, \bibinfo {author} {\bibfnamefont {S.-S.}\ \bibnamefont {Zhang}}, \bibinfo {author} {\bibfnamefont {C.~D.}\ \bibnamefont {Batista}}, \bibinfo {author} {\bibfnamefont {N.}~\bibnamefont {Harrison}}, \bibinfo {author} {\bibfnamefont {R.~D.}\ \bibnamefont {McDonald}}, \bibinfo {author} {\bibfnamefont {L.~E.}\ \bibnamefont {Winter}}, \bibinfo {author} {\bibfnamefont {B.}~\bibnamefont {Ramshaw}}, \bibinfo {author} {\bibfnamefont {M.~K.}\ \bibnamefont {Chan}}, \bibinfo {author} {\bibfnamefont {F.~F.}\ \bibnamefont {Balakirev}}, \emph {et~al.},\ }\bibfield  {title} {\bibinfo {title} {Emergent magnetic anisotropy in the cubic heavy-fermion metal cein$\textsubscript{3}$},\ }\href@noop {} {\bibfield  {journal} {\bibinfo  {journal} {npj Quantum Materials}\ }\textbf {\bibinfo {volume} {2}},\ \bibinfo {pages} {46} (\bibinfo {year} {2017}{\natexlab{b}})}\BibitemShut {NoStop}%
\bibitem [{\citenamefont {Simeth}\ \emph {et~al.}(2023)\citenamefont {Simeth}, \citenamefont {Wang}, \citenamefont {Ghioldi}, \citenamefont {Fobes}, \citenamefont {Podlesnyak}, \citenamefont {Sung}, \citenamefont {Bauer}, \citenamefont {Lass}, \citenamefont {Flury}, \citenamefont {Vonka} \emph {et~al.}}]{simeth2023}%
  \BibitemOpen
  \bibfield  {author} {\bibinfo {author} {\bibfnamefont {W.}~\bibnamefont {Simeth}}, \bibinfo {author} {\bibfnamefont {Z.}~\bibnamefont {Wang}}, \bibinfo {author} {\bibfnamefont {E.}~\bibnamefont {Ghioldi}}, \bibinfo {author} {\bibfnamefont {D.~M.}\ \bibnamefont {Fobes}}, \bibinfo {author} {\bibfnamefont {A.}~\bibnamefont {Podlesnyak}}, \bibinfo {author} {\bibfnamefont {N.}~\bibnamefont {Sung}}, \bibinfo {author} {\bibfnamefont {E.~D.}\ \bibnamefont {Bauer}}, \bibinfo {author} {\bibfnamefont {J.}~\bibnamefont {Lass}}, \bibinfo {author} {\bibfnamefont {S.}~\bibnamefont {Flury}}, \bibinfo {author} {\bibfnamefont {J.}~\bibnamefont {Vonka}}, \emph {et~al.},\ }\bibfield  {title} {\bibinfo {title} {A microscopic kondo lattice model for the heavy fermion antiferromagnet cein3},\ }\href@noop {} {\bibfield  {journal} {\bibinfo  {journal} {Nature communications}\ }\textbf {\bibinfo {volume} {14}},\ \bibinfo {pages} {8239} (\bibinfo {year} {2023})}\BibitemShut {NoStop}%
\bibitem [{\citenamefont {Rosa}\ \emph {et~al.}(2014)\citenamefont {Rosa}, \citenamefont {{de Oliveira}}, \citenamefont {{de Jesus}}, \citenamefont {Moura}, \citenamefont {Adriano}, \citenamefont {Iwamoto}, \citenamefont {Garitezi}, \citenamefont {Granado}, \citenamefont {Saleta}, \citenamefont {Pirota},\ and\ \citenamefont {Pagliuso}}]{ROSA201414}%
  \BibitemOpen
  \bibfield  {author} {\bibinfo {author} {\bibfnamefont {P.}~\bibnamefont {Rosa}}, \bibinfo {author} {\bibfnamefont {L.}~\bibnamefont {{de Oliveira}}}, \bibinfo {author} {\bibfnamefont {C.}~\bibnamefont {{de Jesus}}}, \bibinfo {author} {\bibfnamefont {K.}~\bibnamefont {Moura}}, \bibinfo {author} {\bibfnamefont {C.}~\bibnamefont {Adriano}}, \bibinfo {author} {\bibfnamefont {W.}~\bibnamefont {Iwamoto}}, \bibinfo {author} {\bibfnamefont {T.}~\bibnamefont {Garitezi}}, \bibinfo {author} {\bibfnamefont {E.}~\bibnamefont {Granado}}, \bibinfo {author} {\bibfnamefont {M.}~\bibnamefont {Saleta}}, \bibinfo {author} {\bibfnamefont {K.}~\bibnamefont {Pirota}},\ and\ \bibinfo {author} {\bibfnamefont {P.}~\bibnamefont {Pagliuso}},\ }\bibfield  {title} {\bibinfo {title} {Exploring the effects of dimensionality on the magnetic properties of intermetallic nanowires},\ }\href {https://doi.org/https://doi.org/10.1016/j.ssc.2014.04.013} {\bibfield  {journal} {\bibinfo  {journal} {Solid State Communications}\ }\textbf {\bibinfo
  {volume} {191}},\ \bibinfo {pages} {14} (\bibinfo {year} {2014})}\BibitemShut {NoStop}%
\bibitem [{\citenamefont {Carvalho}\ \emph {et~al.}(2022)\citenamefont {Carvalho}, \citenamefont {Freitas}, \citenamefont {Souza}, \citenamefont {Campanelli}, \citenamefont {Pizzi}, \citenamefont {Mercena}, \citenamefont {Puydinger~dos Santos}, \citenamefont {B\'eron}, \citenamefont {Rosa}, \citenamefont {Pirota},\ and\ \citenamefont {Pagliuso}}]{Carvalho_2022}%
  \BibitemOpen
  \bibfield  {author} {\bibinfo {author} {\bibfnamefont {M.~H.}\ \bibnamefont {Carvalho}}, \bibinfo {author} {\bibfnamefont {G.~S.}\ \bibnamefont {Freitas}}, \bibinfo {author} {\bibfnamefont {J.~C.}\ \bibnamefont {Souza}}, \bibinfo {author} {\bibfnamefont {R.~B.}\ \bibnamefont {Campanelli}}, \bibinfo {author} {\bibfnamefont {H.~B.}\ \bibnamefont {Pizzi}}, \bibinfo {author} {\bibfnamefont {S.~G.}\ \bibnamefont {Mercena}}, \bibinfo {author} {\bibfnamefont {M.}~\bibnamefont {Puydinger~dos Santos}}, \bibinfo {author} {\bibfnamefont {F.}~\bibnamefont {B\'eron}}, \bibinfo {author} {\bibfnamefont {P.~F.~S.}\ \bibnamefont {Rosa}}, \bibinfo {author} {\bibfnamefont {K.}~\bibnamefont {Pirota}},\ and\ \bibinfo {author} {\bibfnamefont {P.~G.}\ \bibnamefont {Pagliuso}},\ }\bibfield  {title} {\bibinfo {title} {Possible routes for the synthesis of nanowires of intermetallic compounds: The case of cein3},\ }\href {https://doi.org/10.1088/1742-6596/2164/1/012041} {\bibfield  {journal} {\bibinfo  {journal} {Journal of Physics:
  Conference Series}\ }\textbf {\bibinfo {volume} {2164}},\ \bibinfo {pages} {012041} (\bibinfo {year} {2022})}\BibitemShut {NoStop}%
\bibitem [{\citenamefont {Tang}\ \emph {et~al.}(2021)\citenamefont {Tang}, \citenamefont {Tang}, \citenamefont {Wu}, \citenamefont {Chen}, \citenamefont {Uzuhashi}, \citenamefont {Ohkubo},\ and\ \citenamefont {Qin}}]{tang2021}%
  \BibitemOpen
  \bibfield  {author} {\bibinfo {author} {\bibfnamefont {S.}~\bibnamefont {Tang}}, \bibinfo {author} {\bibfnamefont {J.}~\bibnamefont {Tang}}, \bibinfo {author} {\bibfnamefont {Y.}~\bibnamefont {Wu}}, \bibinfo {author} {\bibfnamefont {Y.-H.}\ \bibnamefont {Chen}}, \bibinfo {author} {\bibfnamefont {J.}~\bibnamefont {Uzuhashi}}, \bibinfo {author} {\bibfnamefont {T.}~\bibnamefont {Ohkubo}},\ and\ \bibinfo {author} {\bibfnamefont {L.-C.}\ \bibnamefont {Qin}},\ }\bibfield  {title} {\bibinfo {title} {Stable field-emission from a ceb 6 nanoneedle point electron source},\ }\href@noop {} {\bibfield  {journal} {\bibinfo  {journal} {Nanoscale}\ }\textbf {\bibinfo {volume} {13}},\ \bibinfo {pages} {17156} (\bibinfo {year} {2021})}\BibitemShut {NoStop}%
\bibitem [{\citenamefont {Zhang}\ \emph {et~al.}(2022)\citenamefont {Zhang}, \citenamefont {Jimbo}, \citenamefont {Niwata}, \citenamefont {Ikeda}, \citenamefont {Yasuhara}, \citenamefont {Ovidiu}, \citenamefont {Kimoto}, \citenamefont {Kasaya}, \citenamefont {Miyazaki}, \citenamefont {Tsujii} \emph {et~al.}}]{zhang2022}%
  \BibitemOpen
  \bibfield  {author} {\bibinfo {author} {\bibfnamefont {H.}~\bibnamefont {Zhang}}, \bibinfo {author} {\bibfnamefont {Y.}~\bibnamefont {Jimbo}}, \bibinfo {author} {\bibfnamefont {A.}~\bibnamefont {Niwata}}, \bibinfo {author} {\bibfnamefont {A.}~\bibnamefont {Ikeda}}, \bibinfo {author} {\bibfnamefont {A.}~\bibnamefont {Yasuhara}}, \bibinfo {author} {\bibfnamefont {C.}~\bibnamefont {Ovidiu}}, \bibinfo {author} {\bibfnamefont {K.}~\bibnamefont {Kimoto}}, \bibinfo {author} {\bibfnamefont {T.}~\bibnamefont {Kasaya}}, \bibinfo {author} {\bibfnamefont {H.~T.}\ \bibnamefont {Miyazaki}}, \bibinfo {author} {\bibfnamefont {N.}~\bibnamefont {Tsujii}}, \emph {et~al.},\ }\bibfield  {title} {\bibinfo {title} {High-endurance micro-engineered lab6 nanowire electron source for high-resolution electron microscopy},\ }\href@noop {} {\bibfield  {journal} {\bibinfo  {journal} {Nature nanotechnology}\ }\textbf {\bibinfo {volume} {17}},\ \bibinfo {pages} {21} (\bibinfo {year} {2022})}\BibitemShut {NoStop}%
\bibitem [{\citenamefont {Gou}\ \emph {et~al.}(2025)\citenamefont {Gou}, \citenamefont {Tang}, \citenamefont {Huang}, \citenamefont {Guan}, \citenamefont {Zhan}, \citenamefont {Liang}, \citenamefont {Chen}, \citenamefont {Shen},\ and\ \citenamefont {Deng}}]{gou2025}%
  \BibitemOpen
  \bibfield  {author} {\bibinfo {author} {\bibfnamefont {M.}~\bibnamefont {Gou}}, \bibinfo {author} {\bibfnamefont {S.}~\bibnamefont {Tang}}, \bibinfo {author} {\bibfnamefont {K.}~\bibnamefont {Huang}}, \bibinfo {author} {\bibfnamefont {H.}~\bibnamefont {Guan}}, \bibinfo {author} {\bibfnamefont {R.}~\bibnamefont {Zhan}}, \bibinfo {author} {\bibfnamefont {C.}~\bibnamefont {Liang}}, \bibinfo {author} {\bibfnamefont {C.}~\bibnamefont {Chen}}, \bibinfo {author} {\bibfnamefont {Y.}~\bibnamefont {Shen}},\ and\ \bibinfo {author} {\bibfnamefont {S.}~\bibnamefont {Deng}},\ }\bibfield  {title} {\bibinfo {title} {Ultrabright lab6 nanoneedle field emission electron source with low vacuum compatibility},\ }\href@noop {} {\bibfield  {journal} {\bibinfo  {journal} {Communications Materials}\ }\textbf {\bibinfo {volume} {6}},\ \bibinfo {pages} {133} (\bibinfo {year} {2025})}\BibitemShut {NoStop}%
\bibitem [{\citenamefont {Moura}\ \emph {et~al.}(2016)\citenamefont {Moura}, \citenamefont {De~Oliveira}, \citenamefont {Rosa}, \citenamefont {Jesus}, \citenamefont {Saleta}, \citenamefont {Granado}, \citenamefont {B{\'e}ron}, \citenamefont {Pagliuso},\ and\ \citenamefont {Pirota}}]{moura2016}%
  \BibitemOpen
  \bibfield  {author} {\bibinfo {author} {\bibfnamefont {K.}~\bibnamefont {Moura}}, \bibinfo {author} {\bibfnamefont {L.}~\bibnamefont {De~Oliveira}}, \bibinfo {author} {\bibfnamefont {P.~F.~S.}\ \bibnamefont {Rosa}}, \bibinfo {author} {\bibfnamefont {C.}~\bibnamefont {Jesus}}, \bibinfo {author} {\bibfnamefont {M.}~\bibnamefont {Saleta}}, \bibinfo {author} {\bibfnamefont {E.}~\bibnamefont {Granado}}, \bibinfo {author} {\bibfnamefont {F.}~\bibnamefont {B{\'e}ron}}, \bibinfo {author} {\bibfnamefont {P.}~\bibnamefont {Pagliuso}},\ and\ \bibinfo {author} {\bibfnamefont {K.}~\bibnamefont {Pirota}},\ }\bibfield  {title} {\bibinfo {title} {Dimensionality tuning of the electronic structure in fe3ga4 magnetic materials},\ }\href@noop {} {\bibfield  {journal} {\bibinfo  {journal} {Scientific reports}\ }\textbf {\bibinfo {volume} {6}},\ \bibinfo {pages} {28364} (\bibinfo {year} {2016})}\BibitemShut {NoStop}%
\bibitem [{\citenamefont {Moura}\ \emph {et~al.}(2017)\citenamefont {Moura}, \citenamefont {Pirota}, \citenamefont {B{\'e}ron}, \citenamefont {Jesus}, \citenamefont {Rosa}, \citenamefont {Tobia}, \citenamefont {Pagliuso},\ and\ \citenamefont {Lima}}]{moura2017}%
  \BibitemOpen
  \bibfield  {author} {\bibinfo {author} {\bibfnamefont {K.~O.}\ \bibnamefont {Moura}}, \bibinfo {author} {\bibfnamefont {K.~R.}\ \bibnamefont {Pirota}}, \bibinfo {author} {\bibfnamefont {F.}~\bibnamefont {B{\'e}ron}}, \bibinfo {author} {\bibfnamefont {C.~B.}\ \bibnamefont {Jesus}}, \bibinfo {author} {\bibfnamefont {P.~F.~S.}\ \bibnamefont {Rosa}}, \bibinfo {author} {\bibfnamefont {D.}~\bibnamefont {Tobia}}, \bibinfo {author} {\bibfnamefont {P.}~\bibnamefont {Pagliuso}},\ and\ \bibinfo {author} {\bibfnamefont {O.~d.}\ \bibnamefont {Lima}},\ }\bibfield  {title} {\bibinfo {title} {Superconducting properties in arrays of nanostructured $\beta$-gallium},\ }\href@noop {} {\bibfield  {journal} {\bibinfo  {journal} {Scientific reports}\ }\textbf {\bibinfo {volume} {7}},\ \bibinfo {pages} {15306} (\bibinfo {year} {2017})}\BibitemShut {NoStop}%
\bibitem [{\citenamefont {Alexsandro~dos Santos}\ \emph {et~al.}(2021)\citenamefont {Alexsandro~dos Santos}, \citenamefont {Dos~Santos}, \citenamefont {Campanelli}, \citenamefont {Pagliuso}, \citenamefont {Bettini}, \citenamefont {Pirota},\ and\ \citenamefont {B{\'e}ron}}]{alexsandro2021}%
  \BibitemOpen
  \bibfield  {author} {\bibinfo {author} {\bibfnamefont {E.}~\bibnamefont {Alexsandro~dos Santos}}, \bibinfo {author} {\bibfnamefont {M.~V.~P.}\ \bibnamefont {Dos~Santos}}, \bibinfo {author} {\bibfnamefont {R.~B.}\ \bibnamefont {Campanelli}}, \bibinfo {author} {\bibfnamefont {P.~G.}\ \bibnamefont {Pagliuso}}, \bibinfo {author} {\bibfnamefont {J.}~\bibnamefont {Bettini}}, \bibinfo {author} {\bibfnamefont {K.~R.}\ \bibnamefont {Pirota}},\ and\ \bibinfo {author} {\bibfnamefont {F.}~\bibnamefont {B{\'e}ron}},\ }\bibfield  {title} {\bibinfo {title} {Low-temperature electronic transport of manganese silicide shell-protected single crystal nanowires for nanoelectronics applications},\ }\href@noop {} {\bibfield  {journal} {\bibinfo  {journal} {Nanoscale advances}\ }\textbf {\bibinfo {volume} {3}},\ \bibinfo {pages} {3251} (\bibinfo {year} {2021})}\BibitemShut {NoStop}%
\bibitem [{\citenamefont {Cruz}\ \emph {et~al.}(2022)\citenamefont {Cruz}, \citenamefont {Campanelli}, \citenamefont {Dos~Santos}, \citenamefont {Fabris}, \citenamefont {Bettini}, \citenamefont {Pagliuso},\ and\ \citenamefont {Pirota}}]{cruz2022manganese}%
  \BibitemOpen
  \bibfield  {author} {\bibinfo {author} {\bibfnamefont {A.~S.}\ \bibnamefont {Cruz}}, \bibinfo {author} {\bibfnamefont {R.~B.}\ \bibnamefont {Campanelli}}, \bibinfo {author} {\bibfnamefont {M.~V.~P.}\ \bibnamefont {Dos~Santos}}, \bibinfo {author} {\bibfnamefont {F.}~\bibnamefont {Fabris}}, \bibinfo {author} {\bibfnamefont {J.}~\bibnamefont {Bettini}}, \bibinfo {author} {\bibfnamefont {P.~G.}\ \bibnamefont {Pagliuso}},\ and\ \bibinfo {author} {\bibfnamefont {K.~R.}\ \bibnamefont {Pirota}},\ }\bibfield  {title} {\bibinfo {title} {Manganese silicide nanowires via metallic flux nanonucleation: growth mechanism and temperature-dependent resistivity},\ }\href@noop {} {\bibfield  {journal} {\bibinfo  {journal} {Nanotechnology}\ }\textbf {\bibinfo {volume} {33}},\ \bibinfo {pages} {475704} (\bibinfo {year} {2022})}\BibitemShut {NoStop}%
\bibitem [{\citenamefont {Campanelli}\ \emph {et~al.}(2021)\citenamefont {Campanelli}, \citenamefont {dos Santos}, \citenamefont {da~Cruz}, \citenamefont {Pirota},\ and\ \citenamefont {B\'eron}}]{Raul}%
  \BibitemOpen
  \bibfield  {author} {\bibinfo {author} {\bibfnamefont {R.~B.}\ \bibnamefont {Campanelli}}, \bibinfo {author} {\bibfnamefont {M.~V.~P.}\ \bibnamefont {dos Santos}}, \bibinfo {author} {\bibfnamefont {A.~S.~E.}\ \bibnamefont {da~Cruz}}, \bibinfo {author} {\bibfnamefont {K.~R.}\ \bibnamefont {Pirota}},\ and\ \bibinfo {author} {\bibfnamefont {F.}~\bibnamefont {B\'eron}},\ }\bibfield  {title} {\bibinfo {title} {Highly doped si single crystal nanowires via metallic flux nanonucleation},\ }\href {https://doi.org/10.1109/TNANO.2021.3112905} {\bibfield  {journal} {\bibinfo  {journal} {IEEE Transactions on Nanotechnology}\ }\textbf {\bibinfo {volume} {20}},\ \bibinfo {pages} {739} (\bibinfo {year} {2021})}\BibitemShut {NoStop}%
\bibitem [{\citenamefont {Pirota}\ \emph {et~al.}(2020)\citenamefont {Pirota}, \citenamefont {Moura}, \citenamefont {da~Cruz A. S.~E.}, \citenamefont {Campanelli}, \citenamefont {Paliuso},\ and\ \citenamefont {Beron}}]{PIROTA202061}%
  \BibitemOpen
  \bibfield  {author} {\bibinfo {author} {\bibfnamefont {K.}~\bibnamefont {Pirota}}, \bibinfo {author} {\bibfnamefont {K.}~\bibnamefont {Moura}}, \bibinfo {author} {\bibnamefont {da~Cruz A. S.~E.}}, \bibinfo {author} {\bibfnamefont {R.~B.}\ \bibnamefont {Campanelli}}, \bibinfo {author} {\bibfnamefont {P.~G.}\ \bibnamefont {Paliuso}},\ and\ \bibinfo {author} {\bibfnamefont {F.}~\bibnamefont {Beron}},\ }\href@noop {} {\emph {\bibinfo {title} {Intermetallic nanowires fabricated by metallic flux nanonucleation method (MFNN) - Magnetic Nano- and Microwires (Second Edition)}}},\ \bibinfo {edition} {2nd}\ ed.\ (\bibinfo  {publisher} {Woodhead Publishing},\ \bibinfo {year} {2020})\ pp.\ \bibinfo {pages} {61--84}\BibitemShut {NoStop}%
\bibitem [{\citenamefont {Pirota}\ \emph {et~al.}(2014)\citenamefont {Pirota}, \citenamefont {Beron}, \citenamefont {Oliveira}, \citenamefont {Moura}, \citenamefont {Knobel}, \citenamefont {Paliuso}, \citenamefont {Garifezi}, \citenamefont {de~Jesus}, \citenamefont {Rettori}, \citenamefont {Adriano}, \citenamefont {Rosa}, \citenamefont {Urbano}, \citenamefont {Iwamoto}, \citenamefont {Arzuza},\ and\ \citenamefont {Carvalho}}]{Patente}%
  \BibitemOpen
  \bibfield  {author} {\bibinfo {author} {\bibfnamefont {K.}~\bibnamefont {Pirota}}, \bibinfo {author} {\bibfnamefont {F.}~\bibnamefont {Beron}}, \bibinfo {author} {\bibfnamefont {L.}~\bibnamefont {Oliveira}}, \bibinfo {author} {\bibfnamefont {K.}~\bibnamefont {Moura}}, \bibinfo {author} {\bibfnamefont {M.}~\bibnamefont {Knobel}}, \bibinfo {author} {\bibfnamefont {P.~G.}\ \bibnamefont {Paliuso}}, \bibinfo {author} {\bibfnamefont {T.}~\bibnamefont {Garifezi}}, \bibinfo {author} {\bibfnamefont {C.}~\bibnamefont {de~Jesus}}, \bibinfo {author} {\bibfnamefont {C.}~\bibnamefont {Rettori}}, \bibinfo {author} {\bibfnamefont {C.}~\bibnamefont {Adriano}}, \bibinfo {author} {\bibfnamefont {P.}~\bibnamefont {Rosa}}, \bibinfo {author} {\bibfnamefont {R.~R.}\ \bibnamefont {Urbano}}, \bibinfo {author} {\bibfnamefont {W.}~\bibnamefont {Iwamoto}}, \bibinfo {author} {\bibfnamefont {L.}~\bibnamefont {Arzuza}},\ and\ \bibinfo {author} {\bibfnamefont {P.}~\bibnamefont {Carvalho}},\ }\href@noop {} {\bibinfo {title} {Method for
  producing intermetallic monocrystalline nanowires, and intermetallic monocrystalline nanowires}} (\bibinfo {year} {2014})\BibitemShut {NoStop}%
\bibitem [{\citenamefont {Fisk}\ and\ \citenamefont {Remeika.}(1989)}]{Fisk}%
  \BibitemOpen
  \bibfield  {author} {\bibinfo {author} {\bibfnamefont {Z.}~\bibnamefont {Fisk}}\ and\ \bibinfo {author} {\bibfnamefont {J.~P.}\ \bibnamefont {Remeika.}},\ }\bibfield  {title} {\bibinfo {title} {Growth of single crystals from molten metal fluxes},\ }\href {https://doi.org/10.1016/S0168-1273(89)12005-4} {\bibfield  {journal} {\bibinfo  {journal} {Elsevier Science Publishers}\ }\textbf {\bibinfo {volume} {12}},\ \bibinfo {pages} {chapter 81 edition} (\bibinfo {year} {1989})}\BibitemShut {NoStop}%
\bibitem [{\citenamefont {Lee}\ \emph {et~al.}(2006)\citenamefont {Lee}, \citenamefont {Ji}, \citenamefont {G{\"o}sele},\ and\ \citenamefont {Nielsch}}]{lee2006fast}%
  \BibitemOpen
  \bibfield  {author} {\bibinfo {author} {\bibfnamefont {W.}~\bibnamefont {Lee}}, \bibinfo {author} {\bibfnamefont {R.}~\bibnamefont {Ji}}, \bibinfo {author} {\bibfnamefont {U.}~\bibnamefont {G{\"o}sele}},\ and\ \bibinfo {author} {\bibfnamefont {K.}~\bibnamefont {Nielsch}},\ }\bibfield  {title} {\bibinfo {title} {Fast fabrication of long-range ordered porous alumina membranes by hard anodization},\ }\href@noop {} {\bibfield  {journal} {\bibinfo  {journal} {Nature materials}\ }\textbf {\bibinfo {volume} {5}},\ \bibinfo {pages} {741} (\bibinfo {year} {2006})}\BibitemShut {NoStop}%
\bibitem [{\citenamefont {Moshopoulou}\ \emph {et~al.}(2006)\citenamefont {Moshopoulou}, \citenamefont {Ibberson}, \citenamefont {Sarrao}, \citenamefont {L.}, \citenamefont {Thompson},\ and\ \citenamefont {Fisk}}]{Moshopoulou:ws5027}%
  \BibitemOpen
  \bibfield  {author} {\bibinfo {author} {\bibfnamefont {E.~G.}\ \bibnamefont {Moshopoulou}}, \bibinfo {author} {\bibfnamefont {R.~M.}\ \bibnamefont {Ibberson}}, \bibinfo {author} {\bibnamefont {Sarrao}}, \bibinfo {author} {\bibfnamefont {J.}~\bibnamefont {L.}}, \bibinfo {author} {\bibfnamefont {J.~D.}\ \bibnamefont {Thompson}},\ and\ \bibinfo {author} {\bibfnamefont {Z.}~\bibnamefont {Fisk}},\ }\bibfield  {title} {\bibinfo {title} {Structure of ce${\sb 2}$rhin${\sb 8}$: an example of complementary use of high-resolution neutron powder diffraction and reciprocal-space mapping to study complex materials},\ }\href {https://doi.org/10.1107/S0108768106003314} {\bibfield  {journal} {\bibinfo  {journal} {Acta Crystallographica Section B}\ }\textbf {\bibinfo {volume} {62}},\ \bibinfo {pages} {173} (\bibinfo {year} {2006})}\BibitemShut {NoStop}%
\bibitem [{\citenamefont {Blundell}(2001)}]{blundell2001magnetism}%
  \BibitemOpen
  \bibfield  {author} {\bibinfo {author} {\bibfnamefont {S.}~\bibnamefont {Blundell}},\ }\href@noop {} {\emph {\bibinfo {title} {Magnetism in Condensed Matter}}},\ \bibinfo {edition} {2nd}\ ed.\ (\bibinfo  {publisher} {OUP Oxford},\ \bibinfo {address} {Oxford},\ \bibinfo {year} {2001})\BibitemShut {NoStop}%
\bibitem [{\citenamefont {Duque}\ \emph {et~al.}(2011)\citenamefont {Duque}, \citenamefont {Serrano}, \citenamefont {Garcia}, \citenamefont {Bufaical}, \citenamefont {Ferreira}, \citenamefont {Pagliuso},\ and\ \citenamefont {Miranda}}]{duque2011field}%
  \BibitemOpen
  \bibfield  {author} {\bibinfo {author} {\bibfnamefont {J.}~\bibnamefont {Duque}}, \bibinfo {author} {\bibfnamefont {R.~L.}\ \bibnamefont {Serrano}}, \bibinfo {author} {\bibfnamefont {D.}~\bibnamefont {Garcia}}, \bibinfo {author} {\bibfnamefont {L.}~\bibnamefont {Bufaical}}, \bibinfo {author} {\bibfnamefont {L.}~\bibnamefont {Ferreira}}, \bibinfo {author} {\bibfnamefont {P.}~\bibnamefont {Pagliuso}},\ and\ \bibinfo {author} {\bibfnamefont {E.}~\bibnamefont {Miranda}},\ }\bibfield  {title} {\bibinfo {title} {Field induced phase transitions on ndrhin5 and nd2rhin8 antiferromagnetic compounds},\ }\href@noop {} {\bibfield  {journal} {\bibinfo  {journal} {Journal of magnetism and magnetic materials}\ }\textbf {\bibinfo {volume} {323}},\ \bibinfo {pages} {954} (\bibinfo {year} {2011})}\BibitemShut {NoStop}%
\bibitem [{\citenamefont {Lawrence}\ and\ \citenamefont {Shapiro}(1980)}]{lawrence1980}%
  \BibitemOpen
  \bibfield  {author} {\bibinfo {author} {\bibfnamefont {J.}~\bibnamefont {Lawrence}}\ and\ \bibinfo {author} {\bibfnamefont {S.}~\bibnamefont {Shapiro}},\ }\bibfield  {title} {\bibinfo {title} {Magnetic ordering in the presence of fast spin fluctuations: A neutron scattering study of ce in 3},\ }\href@noop {} {\bibfield  {journal} {\bibinfo  {journal} {Physical Review B}\ }\textbf {\bibinfo {volume} {22}},\ \bibinfo {pages} {4379} (\bibinfo {year} {1980})}\BibitemShut {NoStop}%
\bibitem [{\citenamefont {Knafo}\ \emph {et~al.}(2003)\citenamefont {Knafo}, \citenamefont {Raymond}, \citenamefont {F{\aa}k}, \citenamefont {Lapertot}, \citenamefont {Canfield},\ and\ \citenamefont {Flouquet}}]{knafo2003}%
  \BibitemOpen
  \bibfield  {author} {\bibinfo {author} {\bibfnamefont {W.}~\bibnamefont {Knafo}}, \bibinfo {author} {\bibfnamefont {S.}~\bibnamefont {Raymond}}, \bibinfo {author} {\bibfnamefont {B.}~\bibnamefont {F{\aa}k}}, \bibinfo {author} {\bibfnamefont {G.}~\bibnamefont {Lapertot}}, \bibinfo {author} {\bibfnamefont {P.}~\bibnamefont {Canfield}},\ and\ \bibinfo {author} {\bibfnamefont {J.}~\bibnamefont {Flouquet}},\ }\bibfield  {title} {\bibinfo {title} {Study of low-energy magnetic excitations in single-crystalline cein3by inelastic neutron scattering},\ }\href@noop {} {\bibfield  {journal} {\bibinfo  {journal} {Journal of Physics: Condensed Matter}\ }\textbf {\bibinfo {volume} {15}},\ \bibinfo {pages} {3741} (\bibinfo {year} {2003})}\BibitemShut {NoStop}%
\bibitem [{\citenamefont {Kohori}\ \emph {et~al.}(1999)\citenamefont {Kohori}, \citenamefont {Inoue}, \citenamefont {Kohara}, \citenamefont {Tomka},\ and\ \citenamefont {Riedi}}]{kohori1999115in}%
  \BibitemOpen
  \bibfield  {author} {\bibinfo {author} {\bibfnamefont {Y.}~\bibnamefont {Kohori}}, \bibinfo {author} {\bibfnamefont {Y.}~\bibnamefont {Inoue}}, \bibinfo {author} {\bibfnamefont {T.}~\bibnamefont {Kohara}}, \bibinfo {author} {\bibfnamefont {G.}~\bibnamefont {Tomka}},\ and\ \bibinfo {author} {\bibfnamefont {P.}~\bibnamefont {Riedi}},\ }\bibfield  {title} {\bibinfo {title} {115in nqr study in cein3},\ }\href@noop {} {\bibfield  {journal} {\bibinfo  {journal} {Physica B: Condensed Matter}\ }\textbf {\bibinfo {volume} {259}},\ \bibinfo {pages} {103} (\bibinfo {year} {1999})}\BibitemShut {NoStop}%
\end{thebibliography}%

\end{document}